\documentclass[conference,compsoc]{IEEEtran}
\usepackage{cite}
\usepackage{pifont}
\usepackage{amsmath,amssymb,amsfonts}
\usepackage{algorithmic}
\usepackage{graphicx}
\usepackage{textcomp}
\usepackage{makecell}
\usepackage{multirow}
\usepackage{color}
\usepackage[most]{tcolorbox}
\usepackage[table]{xcolor}
\usepackage{enumitem}
\usepackage{listings}
\usepackage{url} 
\usepackage{array}
\usepackage{booktabs}
\usepackage[hidelinks]{hyperref} 
\def\BibTeX{{\rm B\kern-.05em{\sc i\kern-.025em b}\kern-.08em
    T\kern-.1667em\lower.7ex\hbox{E}\kern-.125emX}}
    
\def\X#1{\ding{\numexpr181+#1}}
\IEEEoverridecommandlockouts
\newcommand{\cmark}{\ding{51}}
\newcommand{\xmark}{\ding{55}}

\makeatletter
\renewcommand{\paragraph}{\@startsection{paragraph}{4}%
  {10pt}%
  {0.5ex plus 0.1ex minus 0.1ex}%
  {-0.1em}%
  {\normalfont\normalsize\bfseries}}
\makeatother
\tcbset{rqbox/.style={
  colback=gray!5!white, colframe=black!30,
  top=4pt, bottom=4pt, left=4pt, right=4pt,
  boxsep=1pt, toptitle=1pt, bottomtitle=1pt
}}

\begin{document}

\title{I Don’t Care How You Hide It, I Care What You Do: Runtime Detection of Malicious ML Models}
\author{
\IEEEauthorblockN{Hala Ali}
\IEEEauthorblockA{
\textit{Virginia Commonwealth University, USA} \\
halaali.cs@gmail.com}
\and
\IEEEauthorblockN{Andrew Case}
\IEEEauthorblockA{
\textit{Volatility Foundation, USA} \\
andrew@dfir.org}

}
\maketitle
\begin{abstract}
Pre-trained machine learning (ML) models shared through public platforms can be weaponized to execute malicious payloads during loading or inference. Existing scanners prove inadequate for detecting malicious models as they rely on static analysis and function denylists, support limited serialization formats, and miss attacks that replace denylisted functions with equivalent alternatives, exploit unsupported loading paths, or abuse legitimate framework APIs.
To address this problem, we present \textsc{ModelShield}, the first runtime behavioral detection system for malicious ML models that attributes observed system-level behavior to specific model operations. It combines application-level instrumentation of ML framework APIs with eBPF-based syscall tracing, tracking model behavior across spawned threads and child processes to separate model-triggered activity from environment noise. Attributed syscall sequences are classified using 195 detection patterns mapped to MITRE ATT\&CK techniques. 
We implemented \textsc{ModelShield} to support 17 ML frameworks and 36 formats. On an independent dataset, it is the only evaluated tool to achieve zero false positives, with false negatives limited to payloads that produce no malicious runtime behavior. It detected all 29 published evasion variants spanning function substitution, loading-path obfuscation, and TensorFlow API abuse, where no existing scanner achieved complete coverage. Across 145{,}000 models scanned from public platforms, it captured 41 malicious models on Hugging Face, including a reverse-shell payload and a payload that performed host reconnaissance and outbound exfiltration. Among these, ModelScan, PickleScan, Fickling, and ModelAudit detected only 29\%, 56\%, 80\%, and 93\%, respectively. Despite runtime overhead, \textsc{ModelShield} achieves a median analysis time of 12.3 seconds, with 90\% of scans completing within 35.5 seconds, making it practical for hub-scale screening.
\end{abstract}


\section{Introduction}
\label{intro}
Pre-trained machine learning (ML) models are commonly shared through public platforms such as Hugging Face \cite{huggingface_hub}, Kaggle \cite{kaggle_models}, ModelScope \cite{modelscope}, TensorFlow Hub \cite{tensorflow_hub}, and Keras Hub \cite{kerashub}. While adopting these models reduces the time, cost, and expertise required to deploy ML systems \cite{hitaj2025you, jiang2023empirical, davis2023reusing}, it introduces security risks. ML models can encapsulate learned parameters, architecture, and computational logic, allowing attackers to embed malicious code that is triggered during loading or inference \cite{hepworth2024securing, casey2024large}.

Existing static scanners rely on limited and inconsistent denylist rules, resulting in varying detection results across tools. For instance, PickleScan \cite{maitre2025picklescan}, which is used in Hugging Face's built-in pickle scanning pipeline \cite{huggingface2024pickle}, focuses on pickle-based serialization formats. ModelScan \cite{protectai2025modelscan} supports additional formats such as H5 and TensorFlow SavedModel, but still relies on predefined lists of unsafe functions, imports, and operators. Recent work has shown that these scanners fail against evasion techniques that replace commonly denylisted functions, such as \texttt{os.system}, \texttt{eval}, and \texttt{exec}, with functionally equivalent alternatives from Python's standard library or common ML dependencies. These techniques also exploit loading paths that scanners do not fully support, including nested archives, compression layers, framework-specific packaging formats such as \texttt{zip}, \texttt{tar}, \texttt{gzip}, \texttt{bz2}, and \texttt{xz}, and craft malformed pickle bytecode or archive structures that trigger parser exceptions in scanners while remaining loadable by the framework \cite{liu2025art}. Beyond pickle-based formats, attackers can also abuse legitimate framework APIs, such as TensorFlow SavedModel operations (e.g., \texttt{FixedLengthRecordDatasetV2} and \texttt{DebugIdentityV3}), to perform file leakage, IP exposure,  code execution, and shell access \cite{zhu2025my}.

Addressing these gaps within the current scanners is challenging. Blocking legitimate framework APIs is impractical as benign models and standard ML workflows depend on them \cite{nabeel2026deep}. Expanding denylists to include broader Python standard-library functions or ML dependencies is also insufficient, since the same function can be benign or malicious depending on its arguments and runtime effects \cite{kellas2025pickleball}. Moreover, extending scanner support to all possible loading paths and container formats imposes a continuous maintenance burden as frameworks evolve. Although sandboxing is recommended for executing untrusted models \cite{tensorflow_security}, a sandboxed trace does not inherently determine whether an observed syscall originated from the model artifact, the ML framework, or the execution environment itself \cite{nambiar2026malicious}.

Therefore, we present \textsc{ModelShield}, the first runtime behavioral detection system for malicious ML models that attributes observed system-level behavior to specific model operations. Unlike existing scanners, which flag models containing suspicious or denylisted callables regardless of runtime effect, \textsc{ModelShield} reports a model as malicious only when it produces an observable harmful action. This complements static analysis, which remains necessary for payloads that never execute, while avoiding false positives from dormant or benign uses of flagged functions. \textsc{ModelShield} combines application-level instrumentation of ML framework APIs with eBPF-based syscall tracing to link each captured syscall to the model loading or inference operation that triggered it. This attribution is preserved across threads and child processes, separating model-triggered behavior from framework and execution-environment noise. The attributed syscall sequences are then classified against a taxonomy of 195 detection patterns mapped to MITRE ATT\&CK~\cite{mitre_attack}. To capture behaviors not covered by the taxonomy, the syscall traces are further analyzed offline by an LLM to identify candidate patterns, which are manually reviewed before inclusion.

We implemented \textsc{ModelShield} to support 17 ML frameworks and 36 formats.
On the PickleBall evaluation dataset \cite{kellas2025pickleball}, \textsc{ModelShield} is the only evaluated tool to achieve zero false positives; its false negatives arise from payloads that never execute or produce no malicious behavior during analysis, not from classification errors. It detected all 29 published evasion variants covering function substitution, loading-path obfuscation, and TensorFlow API abuse, where no static tool achieved complete coverage. On a large-scale scan of 105{,}000 Hugging Face, 30{,}000 ModelScope, and 10{,}000 Kaggle models uploaded between March 25 and May 5, 2026, it detected 41 malicious models on Hugging Face, including two real-world malware samples. The first is a reverse-shell payload that remained undetected for three days after upload, with 108 downloads at detection time; the second performed host reconnaissance and outbound exfiltration. Among these, ModelScan, PickleScan, Fickling, and ModelAudit detected only 29\%, 56\%, 80\%, and 93\%, respectively. DynaHug \cite{nambiar2026malicious}, a recent dynamic approach, could analyze only 12\% as it is limited to PyTorch formats and loading-phase monitoring. Although slower than existing scanners due to runtime analysis overhead, \textsc{ModelShield} achieves a median analysis time of 12.3 seconds per model, with 90\% of scans completing within 35.5 seconds, making it practical for hub-scale screening. 

To summarize, we make the following contributions:
\begin{itemize}[leftmargin=20pt]
\item We introduce \textsc{ModelShield}, the first runtime behavioral detection system that attributes syscall activity to specific model loading and inference operations.

\item We design a dual-layer architecture combining AOP-based framework instrumentation with eBPF syscall tracing, propagating correlation across threads and child processes.

\item We develop a taxonomy of 195 detection patterns across eight attack classes mapped to MITRE ATT\&CK, extended via offline LLM-based discovery.

\item We evaluate \textsc{ModelShield} on the PickleBall evaluation dataset, 29 published evasion variants, and 145{,}000 models across three platforms, detecting 41 malicious models including two real-world malware samples, with 90\% of scans completing within 35.5 seconds.

\end{itemize}

\section{Background \& Threat Model}
\label{background}
\subsection{Model Serialization}
\label{format_landscape}
\textbf{Serialization} converts an ML model into a persistent artifact for storage and distribution. Serialization formats vary across frameworks. For instance, PyTorch uses Python's \texttt{pickle} protocol within a ZIP-based container~\cite{pytorch2022_serialization}, TensorFlow SavedModel encodes computational graphs using Protocol Buffers with separate variable checkpoints~\cite{tensorflow2024savedmodel}, and scikit-learn models are serialized via \texttt{joblib} or \texttt{pickle}~\cite{joblib2024}. Additional formats include Keras HDF5~\cite{tensorflow2024hdf5}, ONNX Protobuf~\cite{onnx2024}, and weight-only formats such as SafeTensors and GGUF.
\textbf{Loading, or deserialization}, reconstructs the model via a framework API call such as \texttt{torch.load()} or \texttt{tf.saved\_model.load()}. These APIs read the artifact, rebuild the model structure, restore the parameters, and return a model ready for inference.
\subsection{Instrumentation Primitives}
\textbf{Aspect-Oriented Programming (AOP)} is a programming paradigm that enables runtime interception of function calls without modifying source code \cite{kiczales1997aspect, zheng2024towards}. A \emph{join point} is a program event that can be intercepted, such as a function call; a \emph{pointcut} specifies which join points to intercept; and \emph{advice} is the additional behavior executed at those points. AOP is well suited for instrumenting ML framework APIs, as it enables attaching contextual metadata at specific function boundaries to attribute observed syscalls to the model operations that triggered them.
\textbf{At the kernel layer}, all interactions with the operating system pass through the system call interface, providing an effective observation point for detecting malicious behavior~\cite{ligh2014art}. Extended Berkeley Packet Filter (eBPF)~\cite{cassagnes2020rise, cilium_bpf} is a Linux kernel subsystem that allows user-space programs to run inside the kernel in response to specific events, such as syscall entry and exit tracepoints, capturing system calls and their arguments with minimal overhead \cite{ghigoff2024ebpf_fim}. eBPF also provides kernel-resident maps for sharing state between user space and kernel probes, and perf event buffers for streaming captured events to user space. Unlike \texttt{ptrace}~\cite{ptrace_man}, eBPF programs are invisible to the monitored process, preventing detection or evasion by malicious payloads.

\subsection{Threat Model}
\textbf{Adversary Goal.} The adversary targets downstream users by distributing malicious model artifacts through public platforms. Once downloaded, the embedded payload executes automatically during model loading or inference.
\textbf{Adversary Capability.} The adversary can publish models on public platforms and knows the limitations of existing platform scanners. Based on this, it can embed malicious payloads within model artifacts and develop or adopt evasion techniques to bypass platform-side detection. It can also verify evasion by testing artifacts against static scanners before publication.
\textbf{Assumptions.}
The payload has no prior access to the victim environment, is triggered only when the model is loaded or executed by the victim, and runs with the privileges of the model-running process. The adversary has no prior root- or kernel-level access to tamper with the host kernel, ML framework, or \textsc{ModelShield}; defending against post-exploitation tampering following privilege escalation is outside the scope of this work.

\section{Related Work} 
\label{related}
\paragraph{Static Analysis Approaches}
Kellas et al.~\cite{kellas2025pickleball} proposed \textit{PickleBall}, which analyzes ML library source code to identify callables required to load a given model class and enforces them as an allowlist at load time. Although it blocks the evaluated malicious payloads, it fails to load 20\% of benign models, highlighting the difficulty of building complete allowlists across ML libraries.
Zhao et al.~\cite{zhao2024models} proposed MalHug, which applies taint analysis and heuristic pattern matching to suspicious code extracted from model files and dataset loading scripts on Hugging Face, identifying 91 malicious models across more than 705K analyzed models. Zhu et al. \cite{zhu2025my} focused on detecting TensorFlow SavedModel API abuse; they proposed parsing the computational graph to identify exploitable operations and inspecting their parameters, such as file paths and network endpoints, for suspicious usage.

\paragraph{Dynamic Analysis Approaches}
Casey et al.~\cite{casey2024large} combined Python-level tracing via \texttt{sys.settrace} with \texttt{strace} to monitor model behavior during loading only, focusing on suspicious syscalls such as \texttt{socket}, \texttt{connect}, \texttt{execve}, and \texttt{chmod}, which were then reviewed manually. However, 84\% of flagged models were false positives because this approach lacks the context to distinguish malicious from benign syscall usage. Ravia~\cite{ravia2025smashing} similarly proposed loading models in a sandbox and flagging individual anomalous syscalls and glibc calls, such as process creation or network connections.  Nambiar et al.~\cite{nambiar2026malicious} proposed DynaHug, which executes models in a Docker sandbox, collects \texttt{strace} logs during deserialization, and trains a one-class SVM on syscall presence and frequency features from benign models grouped by task tag (e.g., \texttt{text-generation}). However, it targets only PyTorch pickle models, monitors only the loading phase, trains on all captured syscalls without distinguishing model-triggered behavior from framework and environment noise, and requires a separately trained classifier for each task category.

None of these approaches identifies the origin of observed syscalls, tracks model behavior across spawned threads and processes,  or provides the semantic context needed to distinguish malicious from benign usage. Their analysis is equivalent to generic sandbox monitoring.

\section{Challenges}
\label{challenges}
Designing a runtime behavioral detection system for ML models introduces five key challenges.

\paragraph{C1. Monitoring and Instrumentation Constraints}
Although eBPF enables low-overhead syscall tracing, its kernel verifier enforces strict constraints on stack size, instruction count, and memory access, limiting in-kernel processing. Model execution can generate thousands of syscalls in short bursts, risking event-buffer overflow. Additionally, malicious payloads may crash or hang the model process, 
requiring isolation to keep the monitoring system running and continuous buffer draining to prevent event loss on abrupt termination.

\paragraph{C2. Syscall Attribution Ambiguity}
A single API call, such as \texttt{torch.load()}, can generate thousands
of syscalls from framework internals, CUDA initialization, third-party packages,
and the dynamic linker. At the kernel level, these syscalls are
indistinguishable, making it difficult to determine which activity is triggered by the model artifact, and which belongs to the surrounding environment.

\paragraph{C3. ML Framework Concurrency}
Modern ML frameworks heavily rely on concurrency, distributing work across
multiple threads and child processes.
For instance, PyTorch supports multiprocessing workers for data loading, while TensorFlow uses thread pools for execution of operations.
However, thread pools reuse threads across unrelated tasks, forked processes receive new identifiers, and Python does not natively carry context into thread-pool workers. These factors make it difficult to track model behavior.

\paragraph{C4. Behavioral Pattern Matching and Design Trade-Offs}
Even after attribution, matching syscall sequences against detection patterns is challenging. Unrelated syscalls from different threads can appear adjacent in a single time-ordered stream and be incorrectly matched as a malicious sequence. However, restricting matching to per-thread traces misses attacks that intentionally span threads or processes. Pattern granularity also involves trade-offs, as specific patterns improve precision at the risk of missing attack variations, whereas general patterns increase coverage at 
the cost of false positives. These difficulties are compounded by the varying lengths of attack sequences, from the short syscall burst of a reverse shell to the long repetitive cycles of ransomware encryption.

\paragraph{C5. Model and Execution Heterogeneity}
ML models span diverse frameworks and serialization formats, each with
different threat surfaces. Some formats prevent code execution by design (e.g., SafeTensors), while others permit arbitrary execution during deserialization (e.g., pickle). Therefore, the same syscall behavior carries different security implications depending on the format. For instance, extensive file I/O during a SafeTensors load is benign, whereas the same behavior during pickle deserialization may indicate payload execution.  Even within a single format, the same syscall behavior can be benign or suspicious depending on context, such as accessing \texttt{/proc/cpuinfo} for hardware detection versus host 
reconnaissance. Heterogeneity also affects input inference, as malicious code may trigger only under specific inputs, yet model metadata is often incomplete or missing.

\section{Methodology} 
\label{methodology}
\begin{figure*}[!tb]
\centering
\includegraphics[width=0.95\textwidth]{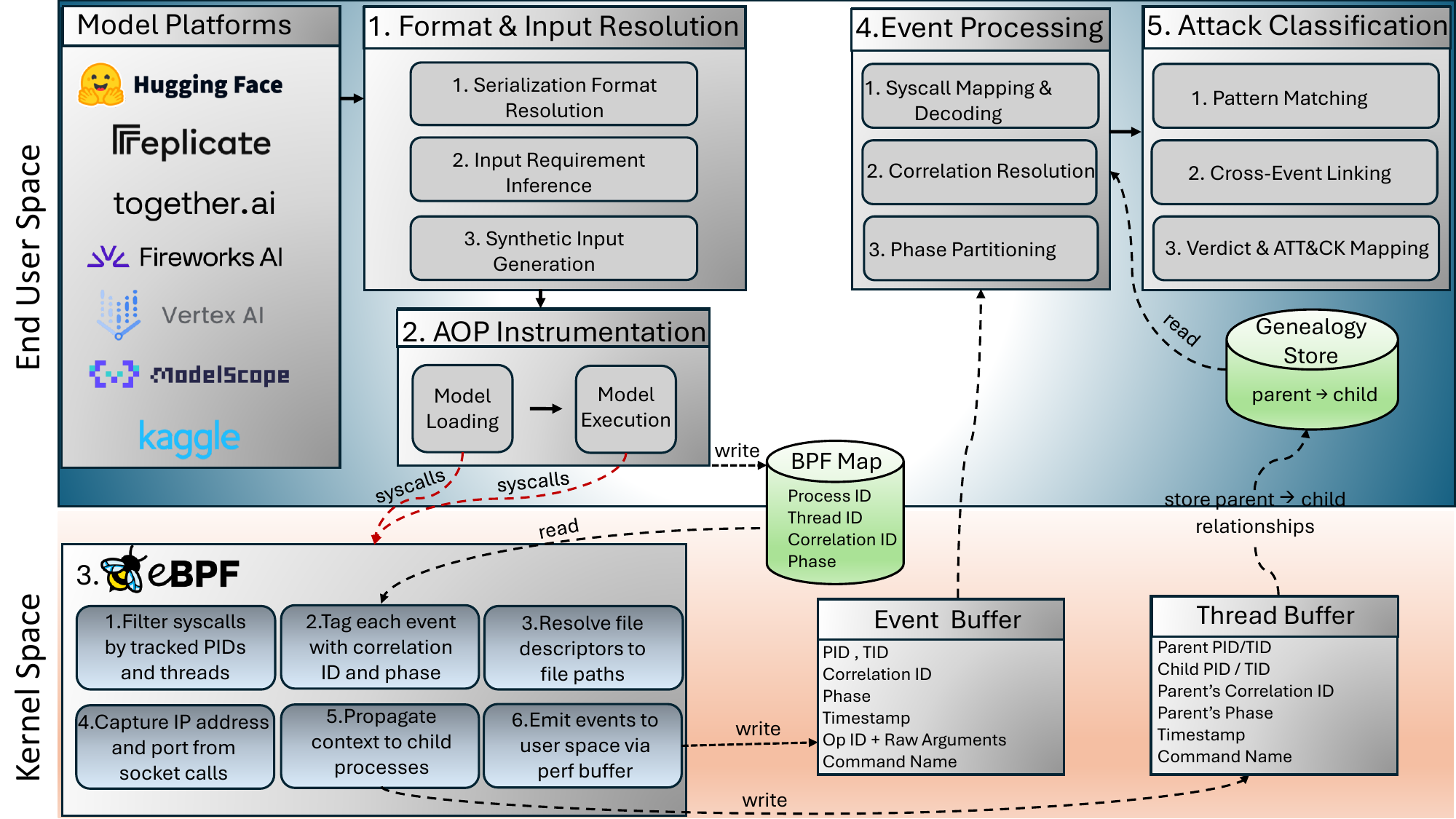}
\caption{Overview of the \textsc{ModelShield} Architecture.}
 \label{meth}
 \vspace{-0.15in}
 \end{figure*}

Figure~\ref{meth} presents the \textsc{ModelShield} architecture, which spans user space and kernel space across five components: \X1~format and input resolution, \X2~application-level instrumentation, \X3~ kernel-level monitoring, \X4~event processing, and \X5~attack behavior classification. 

\subsection{Format and Input Resolution}
\label{format_detection}
Given a model downloaded from a public platform, \textsc{ModelShield} resolves its serialization format and infers the input requirements needed for inference, addressing challenge C5.

\textbf{Format Resolution.}
\textsc{ModelShield} first parses available model metadata, such as configuration files that declare the framework, architecture class, and hyperparameters, to constrain the set of candidate formats. It then matches each model file's extension against 36 supported formats (Table~\ref{format_aop}). When an extension maps to multiple formats, for instance, \texttt{.bin} may correspond to a ZIP-packaged pickle, a TorchScript archive, or a raw pickle stream, \textsc{ModelShield} inspects file headers and magic bytes to resolve the format. When a model spans multiple files, such as a Keras architecture JSON with separate HDF5 weights, \textsc{ModelShield} resolves the dependencies and sequences the loading steps accordingly.
Before loading, structural validation flags files smaller than 1KB as payload-only artifacts, checks for polyglot signatures where a single file matches multiple format types, detects embedded executables such as ELF binaries or shell scripts, and cross-checks declared metadata against the detected framework.

\begin{table*}[!tb]
\centering
\fontsize{8}{9.7}\selectfont
\caption{Supported model formats and AOP interception points for the loading and inference phases.}
\label{format_aop}
\begin{tabular}{p{2.2cm}|p{3.5cm}|p{4.2cm}|p{4.5cm}}
\hline
\textbf{Framework} & \textbf{Format / Extension} & \textbf{Loading Intercepts} & \textbf{Execution Intercepts} \\
\hline
\multirow{2}{*}{PyTorch} & \makecell[l]{.pt, .pth, .bin, .ckpt, .th\\(pickle, TorchScript, package,\\ legacy)} & \makecell[l]{torch.load, torch.jit.load,\\PackageImporter.load\_pickle} & \makecell[l]{\_\_call\_\_, forward} \\
\cline{2-4}
& .pt2 (torch.export) & torch.export.load & \_\_call\_\_ \\
\hline
\multirow{3}{*}{TensorFlow} & SavedModel directory, .pb & tf.saved\_model.load & \makecell[l]{\_\_call\_\_, signatures} \\

\cline{2-4}
& .h5, .keras (Keras) & tf.keras.models.load\_model & \makecell[l]{\_\_call\_\_, predict} \\
\cline{2-4}
& .tflite & tf.lite.Interpreter.\_\_init\_\_ & invoke \\
\hline
\multirow{2}{*}{\makecell[l]{Hugging Face\\Transformers}} & \makecell[l]{.bin, .safetensors} & \makecell[l]{AutoModel.from\_pretrained, \\TFAutoModel.from\_pretrained, \\transformers.pipeline} & \makecell[l]{\_\_call\_\_, forward, \\generate, pipeline.\_\_call\_\_} \\
\cline{2-4}
& .safetensors (standalone) & \makecell[l]{safetensors.torch.load\_file, \\safetensors.numpy.load\_file, \\safetensors.tensorflow.load\_file} & \makecell[l]{delegated to framework \\(e.g., PyTorch, TensorFlow)} \\
\hline
ONNX Runtime & .onnx & \makecell[l]{onnx.load, \\ort.InferenceSession.\_\_init\_\_} & run \\
\hline
\makecell[l]{XGBoost,\\LightGBM,\\CatBoost} & \makecell[l]{.xgb, .bst, .ubj,\\.lgb, .txt, .cbm} & \makecell[l]{xgb.Booster.load\_model,\\lgb.Booster.\_\_init\_\_,\\CatBoost.load\_model} & predict, predict\_proba$^\dagger$ \\
\hline
Scikit-learn & .joblib$^\star$ & joblib.load & \makecell[l]{predict, predict\_proba, \\transform, decision\_function} \\
\hline
JAX / Flax & .msgpack & \makecell[l]{flax.serialization.msgpack\_restore, \\from\_bytes, from\_state\_dict} & \_\_call\_\_, apply \\
\hline
CoreML & .mlmodel, .mlpackage & ct.models.MLModel.\_\_init\_\_ & predict \\
\hline
PaddlePaddle & .pdmodel, .pdiparams & paddle.jit.load, paddle.load & \_\_call\_\_, forward, predict \\
\hline
MXNet & .params, symbol.json & mx.nd.load, mx.sym.load & predict, forward \\
\hline
TensorRT & .engine, .plan, .trt & runtime.deserialize\_cuda\_engine & \makecell[l]{create\_execution\_context, \\execute, execute\_v2, \\execute\_async\_v2$^\S$} \\
\hline
PMML & .pmml & PMMLModel.fromFile & predict \\
\hline
NumPy & .npy, .npz & np.load & load-time only$^\ddagger$ \\
\hline
Pickle & .pkl, .pickle & pickle.load, pickle.loads & (implicit execution on load) \\
\hline
Dill & .dill & \makecell[l]{dill.load, dill.loads, \\dill.load\_module, dill.load\_session} & (implicit execution on load) \\
\hline
\end{tabular}

\vspace{2pt}
\begin{minipage}{0.96\textwidth}
\footnotesize
$^\dagger$\texttt{predict\_proba} is available on sklearn-style wrappers (e.g., \texttt{XGBClassifier}, \texttt{LGBMClassifier}), not on the raw \texttt{Booster} API. Both are instrumented when present.\\
$^\star$Scikit-learn objects serialized as \texttt{.pkl} or \texttt{.pickle} are handled through the generic Pickle.\\
$^\S$TensorRT execution methods are instrumented when an execution context is created; the default inspection path deserializes the engine only.\\
$^\ddagger$Pickle-backed execution occurs only when \texttt{allow\_pickle=True} and the file contains object arrays.
\end{minipage}
 \vspace{-0.15in}
\end{table*}
\textbf{Input Generation.}
\textsc{ModelShield} infers input requirements from multiple sources in priority order. It first extracts metadata fields (e.g., \texttt{model\_type}, \texttt{architectures}, \texttt{image\_size}, and \texttt{vocab\_size}) to determine the expected modality and shape. When metadata is incomplete or absent, it falls back to framework-specific defaults (e.g., tabular feature vectors for gradient-boosting models, tensors shaped to common defaults for ONNX and TFLite), followed by folder-name heuristics matching keywords (e.g.,  \texttt{bert}, \texttt{vit}, or \texttt{whisper}) to their respective modalities. It then generates a synthetic input of the appropriate type and shape, such as a random image tensor for vision models, a token sequence for text models, a waveform tensor for audio models, or a tabular feature vector for classical ML models. For formats where payloads trigger during deserialization (e.g., pickle, dill, and NumPy with object arrays), no input is required.
\subsection{Application-Level Instrumentation} 
Following format resolution and input generation, \textsc{ModelShield} loads and runs the model with application-level instrumentation, addressing challenges C2 and C3.

\textbf{Join Points and Advice.}
\textsc{ModelShield} applies dynamic weaving at the function level via runtime
function interposition. Each framework's loading and inference APIs, listed in Table~\ref{format_aop}, are replaced at import time with instrumented wrappers that preserve the original function signature and logic. After the loading API returns, the wrapper inspects the returned model object via runtime reflection and instruments its callable methods, such as \texttt{generate}, \texttt{decision\_function}, and \texttt{apply}, ensuring coverage of entry points unknown at instrumentation time.

\textbf{Correlation Context.}
When an instrumented framework API (e.g., \texttt{torch.load()}) is called, the AOP wrapper writes a correlation identifier and the current phase (\textsc{Loading} or \textsc{Execution}) into a BPF hash map keyed by the caller's kernel thread identifier. The wrapped call then executes on that same thread, ensuring that every syscall it issues enters the kernel with that thread identifier. At each syscall tracepoint, the eBPF probe in Stage~\X3 looks up the thread identifier in the BPF map and retrieves the correlation context. The phase tag is propagated to the classifier in Stage~\X5, which matches attack patterns separately for loading and inference, as the same syscall sequence may be benign during loading but malicious during inference.

\textbf{Propagation Across Concurrency Boundaries.}
\textsc{ModelShield} propagates correlation context across threads and processes through three complementary mechanisms. First, eBPF probes on \texttt{clone} and \texttt{fork} copy the parent's BPF-map entry to the child at creation time, enabling propagation across kernel-visible boundaries without user-space involvement. Second, Python thread-pool submission is intercepted to initialize each worker's map entry before task execution, preventing the reuse of stale context from previous tasks. Third, a user-space genealogy store records parent-child relationships, enabling retroactive attribution of events whose kernel context was not yet established at capture time.

\textbf{Execution Isolation.}
\textsc{ModelShield} executes the model in a forked process with resource limits on CPU time, memory, file descriptors, and child-process count. The parent process concurrently polls and drains the kernel event buffer, ensuring that events are captured even if the child terminates abnormally due to crashes, \texttt{sys.exit()} calls, or malicious behavior. Continuous draining also prevents buffer overflow during high-volume syscall bursts generated while the model runs.
\subsection{Kernel-Level Monitoring}
While the AOP layer provides semantic context at the application level, the capture of system-level behavior occurs in the kernel via eBPF, addressing challenge C1.

\textbf{Tracepoint Coverage.}
\textsc{ModelShield} attaches eBPF programs to syscall entry and exit tracepoints covering eight categories spanning system, file, network, process, memory, permission, code execution, and kernel module operations. 
On each tracepoint call, the program filters by the set of monitored process identifiers maintained in a BPF map, discarding calls from unrelated processes. The complete list of traced syscalls is provided in Table~\ref{syscall_categories} in the appendix.

\textbf{Event Encoding.}
Given the eBPF verifier constraints (challenge C1), each captured syscall is packaged into a fixed-size record, referred to as an \emph{event}, containing an 8-bit operation identifier that specifies the syscall and the syscall's raw arguments (e.g., file descriptors, flags, pointers). Accordingly, argument decoding, flag interpretation, and semantic labeling are deferred to user space.

\textbf{Event Correlation.}
Each captured event is tagged with the correlation identifier and phase established by the application-level instrumentation. The eBPF program retrieves these values from the shared BPF map using the current thread identifier and stores them in the event record before emission to user space. When correlation metadata is absent, for example, due to delayed context propagation, the event is emitted without attribution and resolved in user space using the multi-level strategy described in Section~\ref{event_processing}. 

\textbf{Semantic Resolution.}
Syscall arguments provide a limited behavioral context. For instance, \texttt{read} takes a file descriptor rather than a file path, and \texttt{connect} takes a pointer to a \texttt{sockaddr} structure rather than its contents. Since these references cannot be dereferenced after the tracepoint returns, \textsc{ModelShield} extracts the underlying values 
in-kernel before event emission. For network syscalls, the eBPF program reads the \texttt{sockaddr} structure and stores the address family, raw IP bytes, and port number in the event record. For file syscalls, it maintains a per-process file-descriptor-to-path map populated at \texttt{openat} and consulted on subsequent \texttt{read}, \texttt{write}, and \texttt{getdents64} calls.

\textbf{Event Streaming.}
Events are emitted to user space through per-CPU perf event buffers, which provide lock-free streaming with minimal overhead. To prevent buffer overflow, the user-space process polls the buffer continuously during model execution. A separate thread-event buffer captures parent-child events emitted by \texttt{clone} and \texttt{fork} probes, recording the parent's correlation context, parent and child identifiers, and the creation timestamp. These events feed the genealogy store that supports retroactive attribution of syscalls from worker threads and child processes.
\subsection{Event Processing}
\label{event_processing}
This stage processes the events received from the eBPF layer through the following steps.

\textbf{Syscall Mapping and Decoding.}
The operation identifier is mapped to its syscall name and to a two-level semantic label consisting of a category (e.g., File, Network, Process) and a sub-operation (e.g., \texttt{File.open}, \texttt{Network.connect}). This allows attack patterns to match semantically equivalent operations regardless of the specific syscall variant. The raw arguments are then decoded into structured fields. Integer flags are mapped to their symbolic constants (e.g., open modes, memory-protection bits, socket families), and network addresses are normalized so that equivalent endpoints, such as IPv4-mapped IPv6 addresses, are represented consistently.

\textbf{Correlation Resolution.}
Once the event is decoded, the next step determines which model operation produced it. Since the eBPF layer emits events only from monitored processes, every event reaching this stage belongs to the model's process tree. The resolution prioritizes the correlation identifier embedded by the eBPF probe. If absent, it falls back to the thread-pool task table, followed by the genealogy store to inherit the parent's correlation context, then a thread-local correlation table maintained at the application level, and finally the correlation context of the model currently under analysis. This ordered fallback ensures that events remain attributed even when context propagation has not yet completed, for example, when a child process emits syscalls before the clone exit probe has propagated the parent's context. Each resolved event is then assigned to the loading-phase or inference-phase stream based on its phase tag.

\subsection{Attack Behavior Classification}
The attributed event streams are then classified against a taxonomy of attack behaviors, addressing challenge C4.

\textbf{Pattern Representation.}
Attack behaviors are defined as patterns of ordered steps, where each step specifies a sub-operation  (e.g., \texttt{File.open}, \texttt{Network.connect})  that must be observed in the event stream. A step may also constrain the target path or flags, such as open modes and socket families, and can be marked as optional. Individual operations that may indicate malicious intent, such as accessing a sensitive file, are not flagged unless a subsequent step produces an observable impact, such as network transmission or code execution. Without such follow-up, flagging the operation would produce false positives on framework activity that incidentally accesses sensitive paths.

\textbf{Pattern Matching.}
The classifier processes each phase-specific event stream in a single linear pass. For each pattern, it maintains a pointer to the current step and advances it when a matching event is observed. Multiple patterns are evaluated simultaneously, each with its own pointer, enabling matching in $O(n \cdot k)$ for $n$ events and $k$ patterns.

\begin{table*}[!htb]
\centering
\fontsize{8}{9.5}\selectfont
\caption{Attack behavior taxonomy and corresponding MITRE ATT\&CK mappings.}
\label{attack_taxonomy}
\begin{tabular}{p{0.6cm}|p{2.8cm}|p{0.6cm}|p{6.6cm}|p{4.5cm}}
\hline
\textbf{ID} & \textbf{Behavior Class} & \textbf{\#} & \textbf{Description} & \textbf{MITRE ATT\&CK} \\
\hline
AB1 & Sensitive Information Theft & 83 &
Reading sensitive files or discovering host, network, and process state, then exfiltrating the collected data over TCP or UDP. &
T1003, T1005, T1016, T1025, T1033, T1039, T1041, T1048, T1049, T1056, T1057, T1059, T1082, T1083, T1550, T1552, T1555, T1560 \\
\hline
AB2 & Backdoor / Persistence & 21 &
Modifying system configuration files or replacing binaries to maintain access across reboots or user sessions. &
T1037, T1053, T1098, T1136, T1542, T1543, T1546, T1547, T1548, T1556, T1574 \\
\hline
AB3 & System Manipulation & 34 &
Deleting, modifying, or tampering with system files, logs, permissions, network settings, or security controls. &
T1036, T1059, T1070, T1098, T1222, T1485, T1548, T1556, T1562, T1565 \\
\hline
AB4 & Malicious Code Execution & 15 &
Spawning processes, staging and executing scripts, or injecting code outside normal model operations. &
T1055, T1059, T1083, T1105, T1106, T1204, T1565, T1567, T1620\\
\hline
AB5 & Cryptojacking & 7 &
Running or deploying cryptocurrency miners, optionally with resource profiling and mining pool communication. &
T1059, T1082, T1105, T1496, T1620 \\
\hline
AB6 & Remote Control & 16 &
Establishing remote access through reverse shells, bind shells, remote command execution, lateral movement, or proxy tunneling. &
T1021, T1036, T1059, T1071, T1090, T1095, T1204 \\
\hline
AB7 & Ransomware & 8 &
Encrypting user files, exfiltrating encryption keys, and inhibiting system recovery. &
T1041, T1083, T1135, T1486, T1490, T1491 \\
\hline
AB8 & Privilege Escalation & 11 &
Elevating process privileges through UID/GID manipulation, SUID abuse, capability changes, kernel module loading, or namespace escape. &
T1068, T1547, T1548, T1611 \\
\hline
\end{tabular}
 \vspace{-0.15in}
\end{table*}

\textbf{Cross-Event Linking.}
Event ordering alone is insufficient to prevent unrelated events from matching a pattern. For instance, a model may read \texttt{/etc/hosts} for DNS resolution and separately open a network connection to download weights; without additional constraints, these unrelated events could falsely match an exfiltration pattern.
To address this, multi-step patterns that span distinct system resources include correlation constraints that require matched events to share a common identifier, such as a file path, file descriptor, or byte count.
\emph{Path correlation} links file events through a shared path. When an event matching a \texttt{File.open} step captures the file path, subsequent \texttt{File.read} or \texttt{File.write} steps advance only if the candidate event operates on the same path. \emph{Socket correlation} links network events through a shared socket descriptor. When an event matching a \texttt{Network.connect} step captures the descriptor, subsequent \texttt{Network.send} steps advance only if the candidate event uses the same descriptor. \emph{Data volume correlation} links file reads to network sends by comparing actual byte counts. \textsc{ModelShield} captures syscall arguments at entry and emits the event at exit, using the return value as the actual bytes transferred. When a \texttt{File.read} event is matched, its byte count is recorded. A subsequent \texttt{Network.send} step advances only if its byte count falls within a plausible range of the recorded value, considering encoding overhead (e.g., Base64 at 1.33$\times$, hex at 2$\times$) and protocol overhead (e.g., JSON or HTTP wrapping).

\textbf{Loop Detection.}
Some attack behaviors produce repetitive syscall cycles that fixed-length patterns cannot capture. \textsc{ModelShield} detects such behavior by tracking repeated operation sequences within the event stream. For instance, it tracks file-operation cycles (\texttt{File.open} $\rightarrow$ \texttt{File.read} $\rightarrow$\texttt{File.write}) across distinct files for ransomware, and sustained send-receive exchanges (\texttt{Network.send} $\rightarrow$ \texttt{Network.recv}) for mining activity.


\textbf{Attack Taxonomy.}
\textsc{ModelShield} defines eight attack behavior classes (AB1--AB8), summarized in Table~\ref{attack_taxonomy} with their descriptions and corresponding MITRE ATT\&CK techniques. Each class contains multiple patterns covering common variants. For instance, AB1 includes 83 patterns spanning TCP and UDP exfiltration of sensitive information (e.g., SSH keys, cloud tokens, browser stores, cryptocurrency wallets, and system configuration), as well as keylogging and command-based discovery. AB4 covers shell and interpreter spawning, script staging and execution, and suspicious subprocess creation. AB6
covers reverse shells, bind shells, remote command execution, lateral movement, and outbound command-and-control connections. Each pattern is an ordered sequence of sub-operations with argument constraints enforced through the correlation mechanism.
To illustrate, the TCP credential-exfiltration pattern in AB1 matches when \texttt{openat} targets a credential file such as \texttt{\texttildelow/.ssh/id\_rsa} in read-only mode, followed by \texttt{read} on the same file descriptor, then \texttt{connect} to an external IP address, and finally \texttt{sendto} on the same socket descriptor. The reverse-shell pattern in AB6 matches when \texttt{socket} creates a TCP endpoint, \texttt{connect} reaches an external host, three \texttt{dup2} calls redirect standard input, output, and error to the connected socket, and \texttt{execve} launches a shell interpreter such as \texttt{/bin/bash}. This taxonomy is derived from reported malicious models and documented attack techniques, and is extended through the LLM-based discovery stage described below. Representative sequences for all eight classes are provided in Table~\ref{ab_sequences} in the appendix.

\textbf{LLM-Based Behavior Discovery.}
The pattern taxonomy is bounded by the behaviors defined at design time. To identify attack behaviors not yet covered, \textsc{ModelShield} includes an offline LLM-based discovery stage applied to models labeled as benign by the classifier. For each model, the loading-phase and execution-phase syscall traces are submitted to an LLM along with a description of expected benign ML runtime behavior and the existing AB1--AB8 category definitions. The LLM is instructed to identify any syscall activity that deviates from the benign baseline and is not covered by the eight attack classes. For each flagged model, the LLM returns a verdict (BENIGN, SUSPICIOUS, or MALICIOUS), confidence level, the relevant syscall subsequence, and MITRE ATT\&CK mappings. Flagged models are then manually reviewed to confirm whether the reported behavior reflects a genuine attack. Confirmed behaviors are added to the taxonomy as new patterns.
\section{Evaluation} 
\label{evaluation}
We evaluate \textsc{ModelShield} through the following research questions:
\begin{itemize}[leftmargin=18pt]
\item \textbf{RQ1.} What is the detection accuracy of \textsc{ModelShield} compared to existing scanners?
\item \textbf{RQ2.} Can \textsc{ModelShield} detect evasion techniques that bypass existing scanners?
\item \textbf{RQ3.} Can \textsc{ModelShield} detect malicious models hosted on public platforms that existing scanners fail to detect?
\item \textbf{RQ4.} How efficient is \textsc{ModelShield} compared to existing scanners at large scale?
\end{itemize}

\textbf{Experimental Setup.}
\label{exp-setup}
We ran all experiments on three virtual machines hosted on a workstation
with an Intel Core i9-13900K processor. Each VM had 4 vCPUs, 16 GB RAM, and ran Ubuntu 20.04 LTS with Linux kernel 5.15, one dedicated to each public platform scan. The application-level instrumentation used dynamic AOP weaving, while the kernel-level monitoring layer was built on the BPF Compiler Collection (BCC). For static scanner comparison, we ran ModelScan \cite{protectai2025modelscan} (v0.8.8), PickleScan \cite{maitre2025picklescan} (v1.0.4), Fickling \cite{fickling} (v0.1.11), and ModelAudit \cite{modelaudit} (v0.2.45). For the LLM-based pattern discovery, we used GPT-5.5. Since building a separate environment per model is infeasible due to time and storage overhead and many models lack dependency declarations, we maintained a pool of five pre-built Python~3.9 virtual environments (see Table~\ref{env-pool} in the appendix), each pinned to a major framework version. The pool covers 99.8\% of the scanned models, distributed as PyTorch (93.0\%), ONNX (3.1\%), classical ML (2.9\%), and TensorFlow/Keras (0.8\%). We excluded the remaining 0.2\% (JAX/Flax, CoreML, PaddlePaddle, MXNet, TensorRT) due to limited prevalence and minimal documented attack activity. For each model, we selected the environment based on platform metadata and file format (see Section~\ref{format_detection}).

\textbf{Evaluation Datasets.}
For detection accuracy (RQ1), we adopted the evaluation dataset published with PickleBall \cite{kellas2025pickleball, kellas2025pickleball_dataset}, which contains 252 benign and 84 malicious models. For evasion robustness (RQ2), we selected 15 gadget substitutions covering denylisted primitives, standard-library wrappers, and ML-library gadgets from Liu et al.~\cite{liu2025art}, along with 10 loading-path obfuscation cases spanning raw pickle, five Joblib compression variants, PyTorch ZIP and TAR containers, NumPy pickle-backed loading, and one EOP bypass. All samples embed an SSH private-key access payload to produce observable credential-access behavior at runtime. We also included all four TensorFlow SavedModel API abuse attacks from Zhu et al.~\cite{zhu2025my, zhu2025tensorabuse_repo}, targeting
IP exposure, file leakage, code execution, and shell access. For large-scale detection and efficiency (RQ3, RQ4), we scanned models uploaded between March~25 and May~5, 2026, on Hugging Face, ModelScope, and Kaggle, yielding 105{,}000, 30{,}000, and 10{,}000 models, respectively. We allocated the largest scanning volume to Hugging Face as it is the most widely adopted platform and the most frequently targeted by both attackers and security researchers.

\subsection{Detection Accuracy Results}
\label{eval_1}
\textsc{ModelShield} produces zero false positives across all 252 benign files of the PickleBall evaluation dataset. On the 84 malicious files, it reports 47 true positives and 37 false negatives. Table~\ref{fn_breakdown} categorizes these false negatives by cause: 1 is a ground-truth mislabel containing no injected payload (\texttt{oceanhacktitude/\allowbreak tinymodel/\allowbreak twitter-roberta-base-\allowbreak sentiment.bin}); 9 arise from environment limitations (5 require packages absent from the runtime environment, all detected after installation, and 4 target Windows-specific APIs); 4 fail before producing any behavior due to syntax errors, undefined variables, or missing target files. The remaining 23 execute successfully but produce no malicious runtime behavior (e.g., printing to standard output or writing to non-sensitive paths). Although all 37 are end-to-end system misses, none represents a classification failure in which \textsc{ModelShield} observed a malicious syscall pattern and failed to flag it. 

\begin{table}[!htb]
\centering
\fontsize{8}{9}\selectfont
\caption{Categorization of the 37 false negatives on the PickleBall evaluation dataset.}
\label{fn_breakdown}
\renewcommand{\arraystretch}{1.2}
\begin{tabular}{p{3.2cm}|r|>{\raggedright\arraybackslash}p{3.8cm}}
\hline
\textbf{Category} & \textbf{\#} & \textbf{Description} \\
\hline
\multicolumn{3}{l}{\cellcolor{gray!15} \textbf{Ground-truth mislabeling (1)}} \\
\hline
Mislabeled sample & 1 & Contains no malicious payload\\
\hline
\multicolumn{3}{l}{\cellcolor{gray!15} \textbf{Environment limitation (9)}} \\
\hline
Missing dependencies & 5 & Requires packages (i.e., \texttt{mlflow}, \texttt{requests}, \texttt{Pillow}, \texttt{cloudpickle}, \texttt{IPython}, \texttt{pydevd}) absent from the environment $^\dagger$ \\

Windows-only payload & 4 & Calls \texttt{nt.system} or \texttt{Calc.exe}\\
\hline
\multicolumn{3}{l}{\cellcolor{gray!15} \textbf{Incomplete execution (4)}} \\
\hline
Defective payload (syntax) & 1 & Places \texttt{return} inside \texttt{eval()}, raising \texttt{SyntaxError} \\ 
Defective payload (runtime) & 1 &  Calls \texttt{print(hello)} with an undefined variable, raising \texttt{NameError} \\
Missing target file & 2 & References a file not included in the model artifact (\texttt{bad\_dataloader.py}, \texttt{exec\_test.py}) \\
\hline
\multicolumn{3}{l}{\cellcolor{gray!15} \textbf{No malicious runtime behavior (23)}} \\
\hline
Print-only payload & 16 & Produces only \texttt{write(stdout)} syscalls \\
Write to non-sensitive paths & 4 & Writes files (e.g., \texttt{p0wn3d.txt}, \texttt{test.txt}, and \texttt{YOUAREHACKED.txt}) using relative paths, and writes \texttt{/tmp/1} under \texttt{/tmp} \\
Bytecode compilation & 1 & Calls \texttt{compile()} to produce bytecode but never passes it to \texttt{exec()}\\
Debugger invocation & 1 & Calls \texttt{pdb.set\_trace()}, blocking on interactive input \\
Benign imports & 1 &  Imports \texttt{\_io.BytesIO} and \texttt{numpy.load} to reconstruct an array from in-memory bytes \\
\hline
\end{tabular}
\\[4pt]
\begin{minipage}{0.95\columnwidth}
\footnotesize
$^\dagger$After installing the required packages and re-executing the five models, \textsc{ModelShield} detected all of them, yielding 52~TP and 32~FN.
\end{minipage}
\end{table}

Table~\ref{pickleball_comparison} compares \textsc{ModelShield} against existing scanners on the PickleBall evaluation dataset. PickleBall \cite{kellas2025pickleball} analyzes library class definitions to generate per-library allowlists of permitted callables, blocking all 84 malicious models whose payloads call callables outside these allowlists. However, it rejects 51 benign models due to namespace mismatches, out-of-class attributes, legacy callables, and secondary pickle loads triggered during inference. ModelScan misses 9 malicious files due to denylist gaps, runtime-only payloads, and multi-\texttt{STOP}-opcode payloads that terminate parsing early. It produces 16 false positives by denylisting \texttt{getattr}, which benign models routinely use. PickleScan misses 5 files due to callables outside its dangerous-imports list, indirect imports through allowed modules, and payloads hidden inside nested pickles, with the same 16 false positives. Fickling fails to parse PyTorch archive formats and crashes on unsupported opcodes, detecting only 2 malicious files while flagging all 252 benign files. ModelAudit misses 2 files, of which one uses a third-party wrapper absent from its denylist. It produces 91 false positives by matching benign weight dictionaries against a CVE-2026-24747 pattern, flagging version-based vulnerabilities, and misreading TorchScript parameter names as dangerous builtins. DynaHug \cite{nambiar2026malicious} flags all 336 files as malicious, as its one-class SVM classifier is trained on syscall profiles from text-generation models. As a result, it flags models from different tasks and libraries as outliers. One file missed by all scanners is the mislabeled RoBERTa model discussed in Table~\ref{fn_breakdown}. PickleBall also blocks this file, as none of its library policies cover the \texttt{transformers} library. Since excluding it would not change the relative ordering of the tools, we report all results against all 84 malicious models.

\begin{table}[!htb]
\centering
\fontsize{8}{8}\selectfont
\caption{Comparison of static and dynamic tools on the PickleBall evaluation dataset.}
\label{pickleball_comparison}
\renewcommand{\arraystretch}{1.3}
\resizebox{\columnwidth}{!}{%
\begin{tabular}{l|rrrrccc}
\hline
\textbf{Tool} & \textbf{TP} & \textbf{TN} & \textbf{FP} & \textbf{FN} & \textbf{Prec.} & \textbf{Recall} & \textbf{F1} \\
\hline
\multicolumn{8}{l}{\cellcolor{gray!15}\textbf{Static Approaches}} \\
\hline
PickleBall             & 84 & 201 &  51 &  0 & 62.22\% & 100.00\%  & 76.71\% \\
ModelScan              & 75 & 236 &  16 &  9 & 82.42\% & 89.29\% & 85.71\% \\
PickleScan             & 79   & 236 &16 &5 & 83.16\%   & 94.05\%  &    88.27\% \\
Fickling               &  2  &  0  & 252  & 82 & 0.79\%  & 2.38\%  &  1.18\% \\
ModelAudit             &   82 &  161 & 91 & 2 & 47.40\%  &   97.62\% &  63.81\% \\
\hline
\multicolumn{8}{l}{\cellcolor{gray!15}\textbf{Dynamic Approaches}} \\
\hline
DynaHug                & 84 &   0 & 252 &  0   & 25.00\% & 100.00\%  & 40.00\% \\
\textsc{ModelShield}   & 47 & 252 &   0 & 37 & 100.00\%  & 55.95\%  & 71.76\% \\

\hline
\end{tabular}
}
\vspace{-0.1in}
\end{table} 
\begin{tcolorbox}[rqbox, title=\textbf{Answer to RQ1}]
\textsc{ModelShield} is the only evaluated tool with zero false positives on the PickleBall evaluation dataset. Its 37 false negatives arise from payloads that either fail to execute due to environment limitations or produce no malicious runtime behavior; none from classification errors.
\end{tcolorbox}
\subsection{Evasion Techniques Detection Results}
\label{eval_2}
\begin{table}[!htb]
\centering
\fontsize{8}{7.5}\selectfont
\caption{Detection results of static scanners and \textsc{ModelShield} against 29 evasion variants.}
\label{evasion_results}
\renewcommand{\arraystretch}{1.3}
\begin{tabular}{p{3.2cm}|p{0.7cm}|p{0.3cm}p{0.3cm}p{0.3cm}p{0.3cm}p{0.3cm}}
\hline
\textbf{Evasion  Variant} & \textbf{Lib} & \textbf{MS} & \textbf{PS} & \textbf{Fi} & \textbf{MA} & \textbf{MS\textsuperscript{$\star$}} \\
\hline
\multicolumn{7}{l}{\cellcolor{gray!15}\textbf{Gadget Substitution~\cite{liu2025art}}} \\
\hline
\texttt{os.system}  & stdlib  &\cmark & \cmark & \cmark& \cmark & \cmark \\
\texttt{pydoc.pipepager} & stdlib   & \xmark  & \cmark  &\cmark &\cmark & \cmark \\
\texttt{pydoc.tempfilepager}   & stdlib  &\xmark  & \cmark  &\cmark & \cmark & \cmark \\
\texttt{doctest.debug\_script} & stdlib  & \xmark & \cmark  & \cmark& \cmark & \cmark \\
\texttt{eval}  & builtin  & \cmark &  \cmark &\cmark & \cmark&  \cmark\\
\texttt{exec}   & builtin  &\cmark  & \cmark  & \cmark& \cmark& \cmark \\
\texttt{np.distutils.\allowbreak exec\_command} & NumPy   &\xmark  & \xmark   & \cmark &\cmark & \cmark \\
\texttt{np.distutils.\allowbreak cpuinfo.getoutput} & NumPy    & \xmark & \xmark &\cmark &\xmark & \cmark \\
\texttt{np.core.\allowbreak \_ufunc\_reconstruct} & NumPy    &\cmark & \cmark  & \cmark&\cmark & \cmark \\
\texttt{sympy.printing.\allowbreak tests.test\_repr.sT} & SymPy    &\xmark  & \xmark & \cmark&\cmark & \cmark \\
\texttt{sympy.sympify}  & SymPy    & \xmark & \xmark &\cmark &\cmark & \cmark \\
\texttt{pandas.\_version.\allowbreak run\_command}  & Pandas  & \cmark & \cmark &\cmark &\cmark & \cmark \\
\texttt{cProfile.run} & stdlib   & \xmark & \cmark & \cmark&\cmark & \cmark \\
\texttt{exec+\_\_import\_\_} & builtin  & \cmark&  \cmark&\cmark &\cmark & \cmark \\
\texttt{subprocess.\allowbreak check\_output} & stdlib   &\cmark & \cmark &\cmark &\cmark  & \cmark \\
\hline
\multicolumn{2}{l|}{\textbf{Gadget Detection}} & 7& 11 & 15&14& 15\\
\hline
\multicolumn{7}{l}{\cellcolor{gray!15}\textbf{Loading Path Obfuscation~\cite{liu2025art}}} \\
\hline
Joblib raw pkl & --  &\cmark & \cmark  & \cmark& \cmark & \cmark \\
Joblib gzip$\to$pkl & --  &\xmark &  \xmark &\xmark &\xmark & \cmark \\
Joblib bz2$\to$pkl & --  & \xmark& \xmark &\xmark & \xmark& \cmark \\
Joblib xz$\to$pkl & --  & \xmark& \xmark &\xmark &\xmark & \cmark \\
Joblib lzma$\to$pkl & --  &\xmark & \xmark &\xmark &\xmark & \cmark \\
Joblib zlib$\to$pkl & --  &\xmark & \xmark &\xmark &\xmark &  \cmark\\
PyTorch zip$\to$pkl & --  & \xmark& \cmark  & \xmark&\cmark & \cmark \\
PyTorch tar$\to$pkl & --  & \xmark& \xmark &\xmark &\cmark & \cmark \\
NumPy npy (pickle) & --  & \cmark& -- & \cmark&\cmark & \cmark \\
EOP-1 STACK\_GLOBAL & --  & \xmark& \cmark &\cmark &\cmark & \cmark \\
\hline
\multicolumn{2}{l|}{\textbf{Path Detection}} &2 & 3 &3 &5& 10\\
\hline
\multicolumn{7}{l}{\cellcolor{gray!15}\textbf{TensorFlow API Abuse~\cite{zhu2025my}}} \\
\hline
IP Exposure & -- &-- & -- & -- &-- & \cmark \\
File Leak & -- &-- & -- & -- &-- & \cmark \\
Code Execution & -- &-- & -- & -- &-- & \cmark \\
Shell Access & -- & -- & -- & -- &-- & \cmark \\
\hline
\multicolumn{2}{l|}{\textbf{TF Detection}} &--  & --  & -- &-- & 4 \\
\hline
\end{tabular}
\\[3pt]
\begin{minipage}{0.95\columnwidth}
\footnotesize
MS\,=\,ModelScan, PS\,=\,PickleScan, Fi\,=\,Fickling, MA\,=\,ModelAudit, MS\textsuperscript{$\star$}\,=\,\textsc{ModelShield}.
\end{minipage}
\vspace{-0.15in}
\end{table}

Table~\ref{evasion_results} compares the static scanners and \textsc{ModelShield} against evasion techniques. ModelScan and PickleScan, limited to their denylists, cover only a narrow set of gadget substitutions and loading paths. Fickling detects all gadget substitutions but cannot parse compressed Joblib or PyTorch container formats. ModelAudit detects 14 of 15 gadget substitutions (hanging on \texttt{numpy.distutils.cpuinfo.\allowbreak getoutput}) but fails on all compressed Joblib variants. 
For TensorFlow API abuse, PickleScan and Fickling do not support the SavedModel format. ModelScan flags File Leak and Shell Access due to \texttt{RpcClient}/\texttt{RpcCall} operations in their graphs, while ModelAudit flags checkpoint operations (\texttt{RestoreV2}, \texttt{SaveV2}, \texttt{MergeV2Checkpoints}) across all four attacks. However, both produce identical findings on benign models containing the same operations; accordingly, their detections are not counted in the table. \textsc{ModelShield} detects all 29 variants, as these techniques modify the delivery mechanism without altering syscall-level behavior.  Detailed evidence is provided in Table~\ref{table_tensorabuse_eval} in the appendix.

\begin{tcolorbox}[rqbox, title=\textbf{Answer to RQ2}]
\textsc{ModelShield} detected all 29 evasion variants across gadget substitution, loading-path obfuscation, and TensorFlow API abuse, where no static scanner achieved complete coverage.
\end{tcolorbox}

\subsection{Large-Scale Detection Results}
\label{eval_3}
\textsc{ModelShield} detected 41 confirmed malicious models during the Hugging Face scan, including two real-world malware samples (Sections~\ref{case_study_lqc} and \ref{case_study_cgruzicka}). However, no malicious models were identified on ModelScope or Kaggle within the evaluated time window. The full set is in Table~\ref{malicious_models_full} in the appendix. Table~\ref{malicious_summary} summarizes the 41 models by format, attack behavior, and execution phase. Pickle is the most targeted format, followed by PyTorch and Joblib.  AB4  is the most common behavior, appearing in 30 models, while AB6  appears in 14, covering outbound connections through standard-library clients, reverse shells, and TLS-based channels. Among pickle-based formats, some payloads execute during deserialization, while others return callable objects whose invocation triggers the malicious behavior. Note that models may exhibit multiple behavior classes, so columns sum to more than 41.

\begin{table}[!htb]
\centering
\fontsize{7.5}{7.5}\selectfont
\caption{Distribution of the 41 malicious Models by Format, Attack Behavior, and Detection Phase.}
\label{malicious_summary}
\renewcommand{\arraystretch}{1.3}
\begin{tabular}{l|rrrrrrr}
\hline
\textbf{Format} & \textbf{\#} & \textbf{AB1} & \textbf{AB3} & \textbf{AB4} & \textbf{AB6} & \textbf{AB8} & \textbf{L / E} \\
\hline
Pickle       & 19 & -- & 4 &  9 &  7 & -- & 12 / 7 \\
PyTorch      & 9 & 1 & -- & 9 &  4 & 1 & 5 / 4 \\
Joblib       &  7 & -- & -- &  6 &  2 & -- & 4 / 3 \\
NumPy        &  4 & -- & -- &  4 &  1 & -- & 1 / 3 \\
Dill         &  1 & -- & -- &  1 & -- & -- & 0 / 1 \\
Scikit-learn &  1 & -- & -- &  1 & -- & -- & 0 / 1 \\
\hline
\textbf{Total} & \textbf{41} & \textbf{1} & \textbf{4} & \textbf{30} & \textbf{14} & \textbf{1} & \textbf{22 / 19} \\
\hline
\end{tabular}
\\[3pt]
\begin{minipage}{0.95\columnwidth}
\footnotesize
AB indicates Attack Behavior, and L/E indicates Loading vs. Execution.
\end{minipage}
\vspace{-0.07in}
\end{table}

Table~\ref{scanner_by_format} compares the static scanners against \textsc{ModelShield} on the 41 malicious models. ModelScan and PickleScan detect only 29\% and 56\%, respectively, as their denylists omit network-capable and utility callables (e.g., \texttt{imaplib.IMAP4}, \texttt{smtplib.SMTP}, \texttt{ftplib.FTP}, \texttt{ssl.get\_server\_certificate}, \texttt{timeit.timeit}) and neither parses compressed Joblib, Dill, or legacy TAR formats. Fickling covers all raw pickle streams, detecting 80\%, yet it does not support container formats, such as compressed Joblib, ZIP-wrapped PyTorch, and NumPy \texttt{.npz} archives. ModelAudit achieves the highest static detection rate at 93\%; however, it hangs indefinitely on three models that reference network-facing modules (e.g., \texttt{imaplib.IMAP4}, \texttt{nntplib.NNTP}).

\begin{table}[!tb]
\centering
\fontsize{8}{8}\selectfont
\caption{Detection results of static scanners and \textsc{ModelShield} against the 41 malicious models.}
\label{scanner_by_format}
\renewcommand{\arraystretch}{1.3}
\begin{tabular}{l|p{0.4cm}|p{0.7cm}p{0.7cm}p{0.7cm}p{0.7cm}p{0.7cm}}
\hline
\textbf{Format} & \textbf{\#} & \textbf{MS} & \textbf{PS} & \textbf{Fi} & \textbf{MA} & \textbf{MS\textsuperscript{$\star$}}\\
\hline
Pickle       & 19 &  3 &  9 & 19 & 16 & 19 \\
PyTorch      & 9 &  2 &  7 &  4 & 9 & 9 \\
Joblib       &  7 &  3 &  4 &  5 &  7 &  7 \\
NumPy        &  4 &  3 &  2 &  3 &  4 &  4 \\
Dill         &  1 &  0 &  0 &  1 &  1 &  1 \\
Scikit-learn &  1 &  1 &  1 &  1 &  1 &  1 \\
\hline
\textbf{Total} & \textbf{41} & \textbf{12} & \textbf{23} & \textbf{33} & \textbf{38} & \textbf{41} \\
\hline
\end{tabular}
\vspace{-0.1in}
\end{table}

To assess recall beyond \textsc{ModelShield}'s own detections, we ran all static scanners on a 10{,}000-model subset. The union of their detections yielded 565 positives, of which manual inspection confirmed 8 as malicious. \textsc{ModelShield} detected 5 of the 8. The 3 misses are end-to-end system failures: 2 models crash the framework parser before payload execution, and 1 reads a sensitive file without any subsequent harmful action. However, none is a classification failure, as no malicious syscall pattern was produced and missed by the classifier. The remaining 557 flags were false positives from denylisted callables used in benign contexts.

We further compare against DynaHug~\cite{nambiar2026malicious, nambiar2026dynahug_repo} on the same 41 models. We found that \texttt{torch.load()}  fails before payload execution on all 32 non-PyTorch files, limiting DynaHug to 9 loadable models. Of those, 4 trigger only during inference, which DynaHug does not monitor, reducing its coverable set to 5 models. Its classifier flags all inputs as anomalous with the same confidence score, including a benign PyTorch model.

\subsubsection{Case Study 1 (Live Reverse Shell)}
\label{case_study_lqc}
\begin{figure*}[!tb] 
\centering
\includegraphics[width=0.95\textwidth]{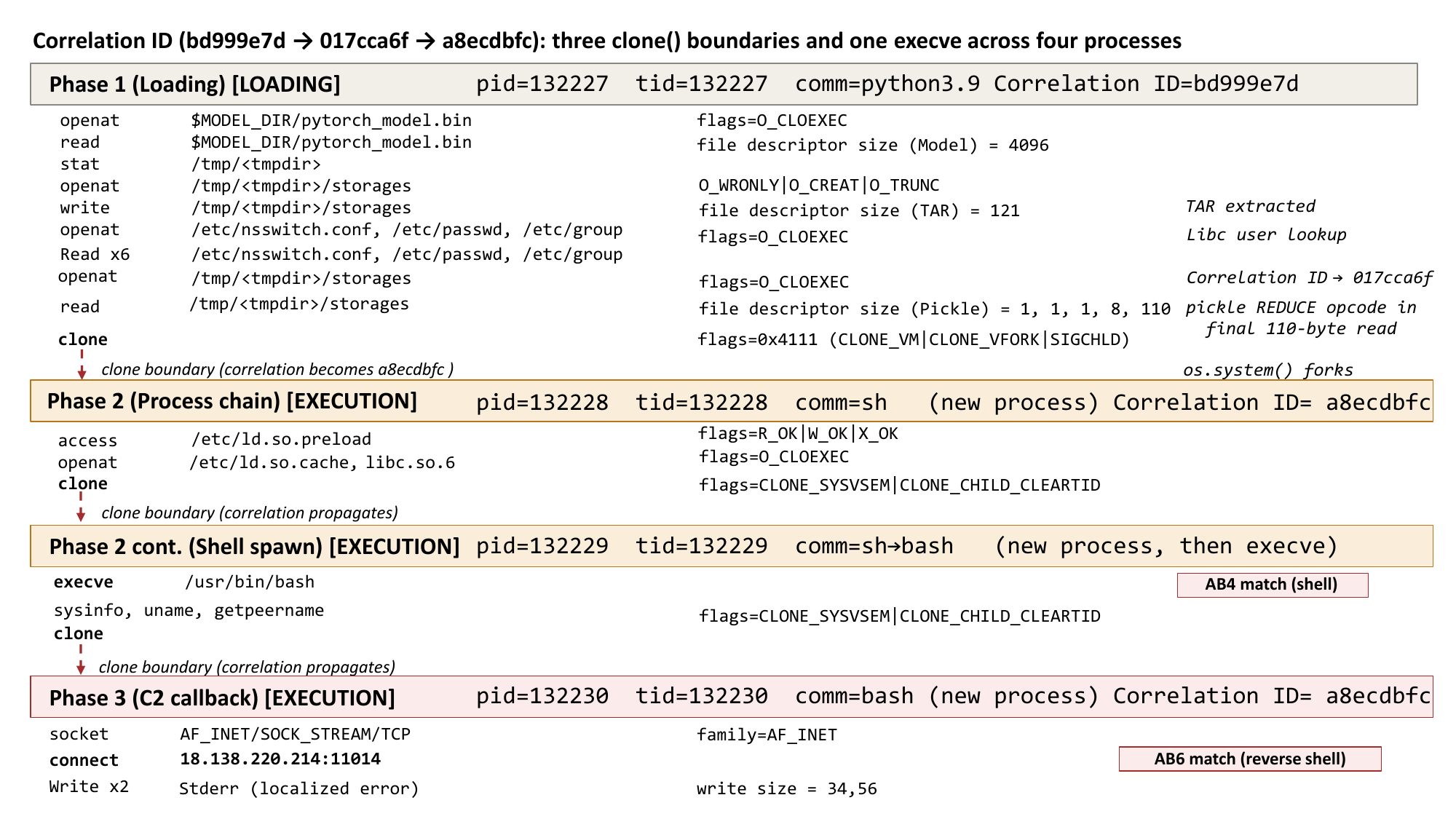}
\caption{Syscall Trace of the Real-World Reverse-Shell Payload Captured by \textsc{ModelShield}.}
 \label{lqct}
 \vspace{-0.15in}
 \end{figure*}
 
\textsc{ModelShield} flagged \texttt{LQCTestingSpace/lqc\_model}, uploaded on March 27, 2026, which had 108 downloads over three days without detection by any of the platform's scanners. The repository presented itself as a Qwen2 model requiring $\sim$300\,MB of weights, yet its \texttt{pytorch\_model.bin} was only 10.2kB. The pickle consisted of a single \texttt{REDUCE} opcode invoking the following command:
\begin{lstlisting}[basicstyle=\ttfamily\footnotesize,breaklines=true,xleftmargin=0.5em,frame=single,framerule=0.3pt,linewidth=\dimexpr\columnwidth-0.5em\relax,columns=flexible]

__import__("os").system(
  "bash -c 'bash -i >& /dev/tcp/18.138.220.214/11014 0>&1'")
\end{lstlisting}

The payload was packaged using PyTorch's legacy tar-based loading path, one of the evasion paths shown by Liu et al.~\cite{liu2025art} to bypass platform scanners. To our knowledge, this is the first documented in-the-wild use of this technique. 
Figure~\ref{lqct} shows how \textsc{ModelShield}'s correlation mechanism tracks the malicious behavior across four processes linked by three \texttt{clone()} boundaries and one \texttt{execve()}. The AOP wrapper tags \texttt{AutoModel.from\_pretrained()} with a correlation identifier (\texttt{bd999e7d}) and phase \textsc{Loading}. When the pickle \texttt{REDUCE} opcode triggers \texttt{os.system()}, the eBPF probe propagates a fresh identifier (\texttt{a8ecdbfc}) to the spawned child, preserving the link to the originating model load. The child then executes \texttt{/usr/bin/bash} (AB4; T1059.004) and connects to \texttt{18.138.220.214:11014} (AB6; T1071).

\subsubsection{Case Study 2 (Host Reconnaissance and Exfiltration)}
\label{case_study_cgruzicka}
\textsc{ModelShield} also flagged \texttt{cgruzicka/\allowbreak test-lora-weights}, a repository containing \texttt{lora\_weights.bin} and \texttt{pytorch\_lora\_weights.bin}, mimicking a legitimate LoRA fine-tuning artifact. During deserialization, the model spawned \texttt{/bin/sh}, executed reconnaissance commands (\texttt{id}, \texttt{whoami}, \texttt{hostname}, \texttt{cat /proc/version}), read \texttt{/etc/passwd} and \texttt{/proc/version}, collected environment variables, and exfiltrated the gathered data over outbound HTTPS connections (port~443) to two IPv4 endpoints (\texttt{178.63.67.153}, \texttt{178.63.67.106}) and two IPv6 endpoints (\texttt{2a01:4f8:\allowbreak 121:114d::2}, \texttt{2a01:4f8:\allowbreak 121:11a5::2}). \textsc{ModelShield} classified this behavior under AB1 (sensitive information theft with exfiltration; T1003.008, T1041), AB4 (shell spawning and reconnaissance; T1059.004, T1082, T1033), and AB6 (outbound connections to attacker-controlled endpoints; T1071).

\begin{tcolorbox}[rqbox, title=\textbf{Answer to RQ3}]
\textsc{ModelShield} detected 41 malicious models on Hugging Face, including models that no evaluated static or dynamic scanner detected.
\end{tcolorbox}
\subsection{Scalability Results}
\label{eval_4}
\begin{table}[!tb]
\centering
\fontsize{7}{7.5}\selectfont
\caption{Per-model timing distribution across evaluated tools.}
\label{rq4_timing}
\renewcommand{\arraystretch}{1.3}
\begin{tabular}{l|p{1.2cm}p{1.2cm}p{1.2cm}p{1.2cm}}
\hline
\textbf{Tool} & \textbf{Mean} & \textbf{Median} & \textbf{P90} & \textbf{Max} \\
 & \textbf{(s)} & \textbf{(s)} & \textbf{(s)} & \textbf{(s)} \\
\hline
\multicolumn{5}{l}{\cellcolor{gray!15}\textbf{ModelShield}} \\
\hline
\textit{Inspector}   & 17.55 & 12.17 & 34.94 & 298.63 \\
\textit{Classifier}  &  0.25 &  0.16 &  0.50 &   5.93 \\
\cline{2-5}
\textit{Total}       & 17.79 & 12.34 & 35.47 & 298.86 \\
\hline
\multicolumn{5}{l}{\cellcolor{gray!15}\textbf{Other Evaluated Scanners}} \\
\hline
ModelScan    &  3.44 &  2.33 &  4.46 & 120.02 \\
PickleScan   &  3.26 &  0.07 &  4.08 & 122.65 \\
Fickling     &  0.21 &  0.06 &  0.08 &  30.03 \\
ModelAudit   &  5.60 &  2.45 & 11.05 & 119.32 \\
DynaHug      &  9.55 &  3.54 & 19.98 & 126.24 \\
\hline
\end{tabular}
\vspace{-0.15in}
\end{table}

\begin{figure}[!tb]
\centering
\includegraphics[width=\columnwidth]{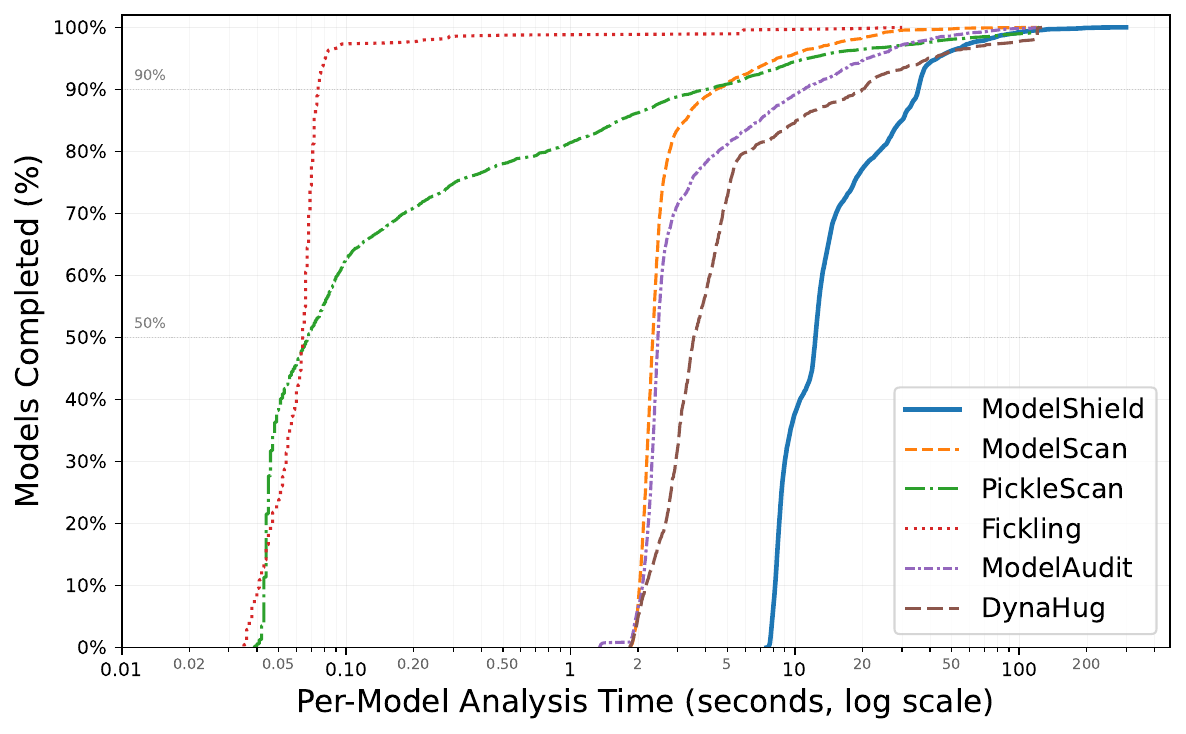}
\caption{CDF of Per-Model Analysis Time.}
\label{rq4_cdf}
\vspace{-0.15in}
\end{figure}

Table~\ref{rq4_timing} presents per-model analysis time across all evaluated tools. \textsc{ModelShield} achieves a median analysis time of 12.34 seconds, comprising inspector and classifier phases; download time is excluded as it depends on platform infrastructure. Each model file is limited to 5 GB and a 300-second inspector timeout. Static scanners are significantly faster, as they analyze model artifacts without loading or executing them. Fickling is the fastest, as it parses pickle bytecode, followed by PickleScan, ModelScan, and ModelAudit. DynaHug achieves a median of 3.54 seconds, as it monitors only the loading phase of PyTorch models. \textsc{ModelShield} introduces additional overhead, at approximately 5$\times$ over ModelScan and ModelAudit and 3.5$\times$ over DynaHug. This reflects the cost of loading and executing each model while tracing and attributing its syscalls across both phases. Figure~\ref{rq4_cdf} shows the cumulative distribution function (CDF), where a point $(x, y)$ indicates that $y$\% of models completed within $x$ seconds. For instance, Fickling scans 90\% of models within 0.08 seconds and ModelScan within 4.46 seconds, while \textsc{ModelShield} requires 35.47 seconds. At this rate, a single \textsc{ModelShield} instance can process approximately 4{,}856 models per day.

\begin{tcolorbox}[rqbox, title=\textbf{Answer to RQ4}]
\textsc{ModelShield} introduces additional overhead compared to the evaluated tools due to its runtime analysis requirements. Despite this, it achieves a median analysis time of 12.3\,s with 90\% of scans complete within 35.5\,s, making it practical for platform-scale scanning.
\end{tcolorbox}

\subsection{LLM-Assisted Discovery Results}
We used GPT-5.5 to analyze 10{,}000 Hugging Face models confirmed as benign by the classifier, covering all major formats. GPT-5.5  flagged \texttt{mathiasthemaster/tabular-playground}, which executes  \texttt{/usr/bin/touch} via a \texttt{posix.system} payload embedded in a pickle \texttt{\_\_reduce\_\_} gadget. After manual confirmation, we added a corresponding pattern to AB4, increasing the taxonomy to 195 patterns. It also produced false positives. For instance, it flagged \texttt{Ievan/attn\_optuna\allowbreak\_06\_04} as malicious with high confidence, interpreting a double \texttt{sys.path} stat walk during \texttt{\_pickle.load} as evidence of embedded \texttt{\_\_import\_\_()} calls.


\section{Discussion}
\label{discussion}
\paragraph{False Positive and False Negative Analysis}
As shown in Section~\ref{eval_1}, \textsc{ModelShield} produces
zero false positives, while its false negatives are limited to payloads
that do not produce observable malicious behavior (see Table~\ref{fn_breakdown}). Static scanners reduce these false negatives by detecting payloads at the artifact level, but at the cost of high false-positive rates. Accordingly, the two approaches are complementary, and deploying both is necessary for robust detection.

\paragraph{Limitations}
As a dynamic analysis system, \textsc{ModelShield} observes only the execution paths reached during analysis. Payloads that depend on specific inputs, environmental conditions, or delayed timers may not activate. Similarly, malformed artifacts that exploit native parser vulnerabilities may crash without producing observable syscall sequences. Malicious payloads may also require specific packages or external files that no runtime analysis system can anticipate; such payloads fail silently without execution. Moreover, the runtime environment targets Linux; thus, payloads targeting other operating systems remain undetected.
\paragraph{Extended Detection Results}
Beyond the evaluation period, we continued deploying \textsc{ModelShield} on Hugging Face through August 19, 2026, scanning 224{,}164 models and detecting 105 additional malicious models, including two real-world malware samples. The first, \texttt{antarge/minilm-finetuned}, is disguised as a MiniLM variant and spawns shell interpreters during deserialization; \textsc{ModelShield} detected it within one hour of upload, with 126 downloads at detection time. The second, \texttt{afdafdfsd/aaaadsdf}, is disguised as a Llama model, harvests credentials from \texttt{/root/.bash\_history}, performs host reconnaissance, and opens a reverse shell to \texttt{78.17.93.96:9000}. \textsc{ModelShield} detected this model three hours after upload, and Hugging Face disabled access following our disclosure.
\section{Conclusion} 
\label{conclusion}
We presented \textsc{ModelShield}, the first runtime behavioral detection system that attributes observed syscall activity to specific model operations. By combining application-level instrumentation with eBPF-based syscall tracing, \textsc{ModelShield} propagates correlation context across threads and child processes, separating model-triggered behavior from framework and environment noise. It then classifies the attributed syscall sequences against 195 detection patterns mapped to MITRE ATT\&CK techniques. On an independent dataset, \textsc{ModelShield} is the only evaluated tool to achieve zero false positives; false negatives arise from payloads that do not produce observable malicious behavior during analysis, none of which represent classification errors. It also achieved complete detection across the 29 evasion variants that bypass existing static scanners. On 145{,}000 models across three public platforms, \textsc{ModelShield} detected 41 malicious models on Hugging Face, including two real-world malware samples. Despite higher runtime overhead compared to existing scanners, \textsc{ModelShield} achieves a median analysis time of 12.3\,s per model, making it practical for platform-scale screening.

\section{Data Availability} 
\label{availability}
The \textsc{ModelShield} source code, detection taxonomy, evaluation test suites, and reproduction scripts are available at \url{https://github.com/HalaAli198/ModelShield}. The repository also provides full detection details for all detected malicious models and security advisories for the four discovered real-world malware samples.

 \section{Ethical Considerations} 
 \label{Ethical}
We reported all real-world malware samples to the Hugging Face Security Team. We also reported the reverse-shell payload to AWS Trust \& Safety, as its command-and-control server was hosted on AWS infrastructure; Hugging Face removed the model within two days, and AWS took action on the reported endpoint. All analysis was performed in isolated virtual machines containing no sensitive data.

\section{LLM Usage Statement}
LLMs were used for editorial purposes in this manuscript, and all outputs were inspected by the authors to ensure accuracy and originality.
\bibliographystyle{IEEEtran}
\bibliography{references}

@article{casey2024large,
  title={A large-scale exploit instrumentation study of {AI/ML} supply chain attacks in {Hugging Face} models},
  author={Casey, Beatrice and Santos, Joanna and Mirakhorli, Mehdi},
  journal={arXiv preprint arXiv:2410.04490},
  year={2024}
}

@inproceedings{liu2025art,
  title={The Art of Hide and Seek: Making {Pickle-Based} Model Supply Chain Poisoning Stealthy Again},
  author={Liu, Tong and Meng, Guozhu and Zhou, Peng and Deng, Zizhuang and Yao, Shuaiyin and Chen, Kai},
  booktitle={Proceedings of the 35th USENIX Security Symposium (USENIX Security 26)},
  pages={161--180},
  year={2026}
}

@inproceedings{zhu2025my,
  title={My model is malware to you: Transforming {AI} models into malware by abusing {TensorFlow} {APIs}},
  author={Zhu, Ruofan and Chen, Ganhao and Shen, Wenbo and Xie, Xiaofei and Chang, Rui},
  booktitle={Proceedings of the 2025 IEEE Symposium on Security and Privacy (SP)},
  pages={486--503},
  year={2025}
}

@inproceedings{kellas2025pickleball,
  title={{PickleBall}: Secure Deserialization of Pickle-based Machine Learning Models},
  author={Kellas, Andreas D and Christou, Neophytos and Jiang, Wenxin and Li, Penghui and Simon, Laurent and David, Yaniv and Kemerlis, Vasileios P and Davis, James C and Yang, Junfeng},
  booktitle={Proceedings of the 2025 ACM SIGSAC Conference on Computer and Communications Security (CCS)},
  pages={3341--3355},
  year={2025}
}

@misc{kellas2025pickleball_dataset,
  author       = {Kellas, Andreas D. and Christou, Neophytos and Jiang, Wenxin and Li, Penghui and Simon, Laurent and David, Yaniv and Kemerlis, Vasileios P. and Davis, James C. and Yang, Junfeng},
  title        = {{PickleBall}: Secure Deserialization of Pickle-based Machine Learning Models -- Artifact},
  year         = {2025},
  doi          = {10.5281/zenodo.16974644},
  url          = {https://doi.org/10.5281/zenodo.16974644},
  note         = {Accessed: 2026-08-19}
}

@inproceedings{zhao2024models,
  title={Models are codes: Towards measuring malicious code poisoning attacks on pre-trained model hubs},
  author={Zhao, Jian and Wang, Shenao and Zhao, Yanjie and Hou, Xinyi and Wang, Kailong and Gao, Peiming and Zhang, Yuanchao and Wei, Chen and Wang, Haoyu},
  booktitle={Proceedings of the 39th IEEE/ACM International Conference on Automated Software Engineering (ASE)},
  pages={2087--2098},
  year={2024}
}

@misc{ravia2025smashing,
  author       = {Itay Ravia},
  title        = {Smashing Model Scanners: Advanced Bypass Techniques and a Novel Detection Approach},
  howpublished = {Black Hat USA 2025},
  year         = {2025},
  url          = {https://blackhat.com/us-25/briefings/schedule/},
  note         = {Accessed: 2026-03-31}
}

@misc{maitre2025picklescan,
  author = {Matthieu Maitre},
  title  = {picklescan},
  year   = {2025},
  url    = {https://github.com/mmaitre314/picklescan},
  note   = {Accessed: 2026-03-31}
}

@misc{protectai2025modelscan,
  author = {{ProtectAI}},
  title  = {modelscan},
  year   = {2025},
  url    = {https://github.com/protectai/modelscan},
  note   = {Accessed: 2026-03-31}
}

@misc{huggingface2024pickle,
  author = {{Hugging Face}},
  title  = {Pickle scanning},
  year   = {2024},
  url    = {https://huggingface.co/docs/hub/security-pickle},
  note   = {Accessed: 2026-03-31}
}

@misc{fickling,
  author       = {{Trail of Bits}},
  title={Fickling: A {Python} pickle decompiler and static analyzer},
  year         = {2021},
  howpublished = {\url{https://github.com/trailofbits/fickling}},
  note         = {Accessed: 2026-04-12}
}

@misc{modelaudit,
  author       = {{Booz Allen Hamilton}},
  title        = {{ModelAudit}},
  year         = {2025},
  howpublished = {\url{https://github.com/bah-insignia/modelaudit}},
  note         = {Accessed: 2026-03-31}
}

@inproceedings{kiczales1997aspect,
  title={Aspect-oriented programming},
  author={Kiczales, Gregor and Lamping, John and Mendhekar, Anurag and Maeda, Chris and Lopes, Cristina and Loingtier, Jean-Marc and Irwin, John},
  booktitle={Proceedings of the 11th European Conference on Object-Oriented Programming (ECOOP)},
  pages={220--242},
  year={1997}
}

@misc{pytorch2022_serialization,
  title={{PyTorch} Serialization Documentation},
  author       = {{PyTorch Contributors}},
  year         = {2022},
  howpublished = {\url{https://pytorch.org/docs/stable/notes/serialization.html}},
  note         = {Accessed: 2026-04-01}
}

@misc{tensorflow2024savedmodel,
  title={{TensorFlow} {SavedModel} Guide},
  author       = {{TensorFlow Team}},
  year         = {2024},
  howpublished = {\url{https://www.tensorflow.org/guide/saved_model}},
  note         = {Accessed: 2026-04-01}
}

@misc{tensorflow2024hdf5,
  title={Save and Load Models in {Keras}},
  author       = {{TensorFlow Team}},
  year         = {2024},
  howpublished = {\url{https://www.tensorflow.org/tutorials/keras/save_and_load}},
  note         = {Accessed: 2026-04-01}
}

@misc{joblib2024,
  title        = {Joblib: Persistence},
  author       = {{Joblib Developers}},
  year         = {2024},
  howpublished = {\url{https://joblib.readthedocs.io/en/latest/persistence.html}},
  note         = {Accessed: 2026-04-01}
}

@misc{onnx2024,
  title={{ONNX}: Open Neural Network Exchange},
  author       = {{ONNX Contributors}},
  year         = {2024},
  howpublished = {\url{https://onnx.ai/onnx/}},
  note         = {Accessed: 2026-04-01}
}

@inproceedings{zheng2024towards,
  title={Towards robust detection of open source software supply chain poisoning attacks in industry environments},
  author={Zheng, Xinyi and Wei, Chen and Wang, Shenao and Zhao, Yanjie and Gao, Peiming and Zhang, Yuanchao and Wang, Kailong and Wang, Haoyu},
  booktitle={Proceedings of the 39th IEEE/ACM International Conference on Automated Software Engineering (ASE)},
  pages={1990--2001},
  year={2024}
}

@inproceedings{cassagnes2020rise,
  title={The rise of {eBPF} for non-intrusive performance monitoring},
  author={Cassagnes, Cyril and Trestioreanu, Lucian and Joly, Clement and State, Radu},
  booktitle={Proceedings of the 2020 IEEE/IFIP Network Operations and Management Symposium (NOMS)},
  pages={1--7},
  year={2020}
}

@misc{cilium_bpf,
  author = {{Cilium}},
  title={{BPF} and {XDP} Reference Guide},
  url    = {https://docs.cilium.io/en/stable/reference-guides/bpf/},
  note   = {Accessed: 2026-04-02}
}

@misc{ghigoff2024ebpf_fim,
  author = {Yoann Ghigoff and Jonathan Ribas and Sylvain Afchain and Sylvain Baubeau and Guillaume Fournier},
  title={Workload Protection with {eBPF}-based File Integrity Monitoring},
  year   = {2024},
  url    = {https://www.datadoghq.com/blog/engineering/workload-protection-ebpf-fim/},
  note   = {Accessed: 2026-04-02}
}

@book{ligh2014art,
  title={The Art of Memory Forensics: Detecting Malware and Threats in Windows, Linux, and Mac Memory},
  author={Ligh, Michael Hale and Case, Andrew and Levy, Jamie and Walters, Aaron},
  year={2014},
  publisher={John Wiley \& Sons}
}

@misc{ptrace_man,
  author = {{Linux man-pages project}},
  title={ptrace(2) - {Linux} manual page},
  url    = {https://man7.org/linux/man-pages/man2/ptrace.2.html},
  note   = {Accessed: 2026-04-02}
}

@inproceedings{jiang2023empirical,
  title={An empirical study of pre-trained model reuse in the {Hugging Face}  deep learning model registry},
  author={Jiang, Wenxin and Synovic, Nicholas and Hyatt, Matt and Schorlemmer, Taylor R and Sethi, Rohan and Lu, Yung-Hsiang and Thiruvathukal, George K and Davis, James C},
  booktitle={Proceedings of the 2023 IEEE/ACM 45th International Conference on Software Engineering (ICSE)},
  pages={2463--2475},
  year={2023}
}

@article{hitaj2025you,
  title={Do you trust your model? {E}merging malware threats in the deep learning ecosystem},
  author={Hitaj, Dorjan and Pagnotta, Giulio and De Gaspari, Fabio and Ruko, Sediola and Hitaj, Briland and Mancini, Luigi V and Perez-Cruz, Fernando},
  journal={IEEE Transactions on Dependable and Secure Computing},
  year={2025},
  publisher={IEEE}
}

@inproceedings{davis2023reusing,
  title={Reusing deep learning models: Challenges and directions in software engineering},
  author={Davis, James C and Jajal, Purvish and Jiang, Wenxin and Schorlemmer, Taylor R and Synovic, Nicholas and Thiruvathukal, George K},
  booktitle={Proceedings of the 2023 IEEE John Vincent Atanasoff International Symposium on Modern Computing (JVA)},
  pages={17--30},
  year={2023}
}

@misc{huggingface_hub,
  author       = {{Hugging Face, Inc.}},
  title        = {{Hugging Face Hub Documentation}},
  howpublished = {\url{https://huggingface.co/docs/hub/index}},
  year         = {2026},
  note         = {Accessed: 2026-04-14}
}

@misc{kaggle_models,
  author       = {{Kaggle, Inc.}},
  title        = {{Kaggle Models}},
  howpublished = {\url{https://www.kaggle.com/models}},
  year         = {2026},
  note         = {Accessed: 2026-04-14}
}

@misc{kerashub,
  author       = {Watson, M. and Chollet, F. and Sreepathihalli, D. and Saadat, S. and Sampath, R. and Rasskin, G. and Zhu, S. and Singh, V. and Wood, L. and Tan, Z. and Stenbit, I. and Qian, C. and Bischof, J. and others},
  title        = {{KerasHub}},
  howpublished = {\url{https://github.com/keras-team/keras-hub}},
  year         = {2024}
}

@misc{tensorflow_hub,
  author       = {{Google Research, Brain Team}},
  title        = {{TensorFlow Hub: Reusable Machine Learning Modules}},
  howpublished = {\url{https://www.tensorflow.org/hub}},
  year         = {2026},
  note         = {Accessed: 2026-04-14}
}

@misc{modelscope,
  author       = {{Alibaba DAMO Academy}},
  title        = {{ModelScope}},
  howpublished = {\url{https://modelscope.cn}},
  year         = {2026},
  note         = {Accessed: 2026-04-14}
}

@misc{mitre_attack,
  author = {{MITRE Corporation}},
  title = {{MITRE ATT\&CK\textregistered{} Knowledge Base}},
  year = {2024},
  url = {https://attack.mitre.org/},
  note = {Accessed: 2026-04-14}
}

@article{nabeel2026deep,
  title={Deep Dive into the Abuse of {DL} {APIs} To Create Malicious {AI} Models and How to Detect Them},
  author={Nabeel, Mohamed and Starov, Oleksii},
  journal={arXiv preprint arXiv:2601.04553},
  year={2026}
}

@misc{tensorflow_security,
  title={Using {TensorFlow} Securely},
  author       = {{TensorFlow}},
  howpublished = {\url{https://github.com/tensorflow/tensorflow/security}},
  note         = {Accessed: 2026-05-10}
}

@article{nambiar2026malicious,
  title={Malicious {ML} Model Detection by Learning Dynamic Behaviors},
  author={Nambiar, Sarang and Pradhan, Dhruv and Soremekun, Ezekiel},
  journal={arXiv preprint arXiv:2604.19438},
  year={2026}
}

@misc{nambiar2026dynahug_repo,
  author={Nambiar, Sarang and Pradhan, Dhruv and Soremekun, Ezekiel},
  title={{DynaHug}: Source Code and Artifacts},
  year={2026},
  howpublished={\url{https://github.com/DynaHug-Detector/DynaHug}},
  note={Accessed: 2026-05-25}
}

@misc{zhu2025tensorabuse_repo,
  author       = {Ruofan Zhu and Ganhao Chen and Wenbo Shen and Xiaofei Xie and Rui Chang},
  title        = {{TensorAbuse}: Practical Attacks},
  year         = {2025},
  howpublished = {\url{https://github.com/ZJU-SEC/TensorAbuse}},
  note         = {Accessed: 2026-05-24}
}

@techreport{hepworth2024securing,
  title       = {Securing the {AI} Software Supply Chain},
  author      = {Hepworth, Isaac and Olive, Kara and Dasgupta, Kingshuk 
                 and Le, Michael and Lodato, Mark and Maruseac, Mihai 
                 and Meiklejohn, Sarah and Chaudhuri, Shamik 
                 and Minkus, Tehila},
  institution = {Google},
  year        = {2024}
}


\appendices
\section{Supplementary Tables}
Table~\ref{syscall_categories} lists all syscalls monitored by the eBPF layer, grouped by category. Table~\ref{ab_sequences} shows one representative ordered syscall sequence per sub-behavior for each attack class. Table \ref{env-pool} details the composition of the runtime environment pool. Table~\ref{table_tensorabuse_eval} details the runtime evidence captured for each TensorAbuse attack variant. Table~\ref{malicious_models_full} lists all 41 confirmed malicious models detected on Hugging Face. 

\begin{table}[!htb]
\centering
\fontsize{9.5}{11}\selectfont
\caption{Traced syscalls grouped by category.}
\label{syscall_categories}
\begin{tabular}{l|l}
\hline
\textbf{Category} & \textbf{Syscalls} \\
\hline
\multirow{3}{*}{\textbf{System}} & \makecell[l]{uname, sysinfo, sched\_getaffinity, \\ioctl, getrandom, sethostname, \\setdomainname} \\
\hline
\multirow{10}{*}{\textbf{File}} & \makecell[l]{open, openat, creat, close, \\read, pread, pread64, readv,\\ write, pwrite, pwrite64, writev, \\stat, fstat, newfstatat, statx, \\access, faccessat, getdents64, getcwd, \\rename, renameat, renameat2, \\link, linkat, symlink, symlinkat, \\unlink, unlinkat, rmdir, \\chmod, fchmod, fchmodat, \\chown, fchown, fchownat, lchown, \\mkdir, mkdirat, readlink, readlinkat, \\truncate, ftruncate, fcntl, lseek, \\chdir, fchdir, sendfile, dup, dup2, dup3, \\setxattr, lsetxattr, fsetxattr, \\removexattr, lremovexattr, fremovexattr, \\pipe, pipe2, inotify\_init, inotify\_init1, \\inotify\_add\_watch, inotify\_rm\_watch} \\
\hline
\multirow{4}{*}{\textbf{Network}} & \makecell[l]{socket, bind, listen, accept, accept4, \\connect, sendto, sendmsg, \\recvfrom, recvmsg, \\getsockname, getpeername, \\getsockopt, setsockopt} \\
\hline
\multirow{2}{*}{\textbf{Process}} & \makecell[l]{fork, vfork, clone, clone3, \\prctl, ptrace, setsid, setns, unshare} \\
\hline
\textbf{Memory} & \makecell[l]{mmap, mprotect, memfd\_create} \\
\hline
\multirow{3}{*}{\textbf{Permission}} & \makecell[l]{getuid, geteuid, getgid, getegid, \\getgroups, setuid, setgid, setreuid, \\setresuid, setregid, setresgid, \\setgroups, capset} \\
\hline
\textbf{Code Execution} & \makecell[l]{execve, execveat} \\
\hline
\multirow{2}{*}{\textbf{Kernel Module}} & \makecell[l]{init\_module, finit\_module, \\delete\_module} \\
\hline
\end{tabular}
\end{table}

\begin{table*}[!tb]
\centering
\fontsize{8}{11}\selectfont
\caption{Representative syscall sequences for each attack class. Arrows indicate temporal ordering; parenthesized steps are optional.}
\label{ab_sequences}
\begin{tabular}{p{0.6cm}|p{3.9cm}|p{12.2cm}}
\hline
\textbf{ID} & \textbf{Sub-Behavior} & \textbf{Ordered Sequence} \\
\hline
\multirow{8}{*}{AB1}
 & Credential exfiltration (TCP) &
\texttt{openat}(sensitive file, \texttt{O\_RDONLY}) $\rightarrow$ \texttt{read} $\rightarrow$ (\texttt{socket}) $\rightarrow$ \texttt{connect}(remote) $\rightarrow$ \texttt{sendto} \\
\cline{2-3}
 & Credential exfiltration (UDP) &
\texttt{openat}(sensitive file, \texttt{O\_RDONLY}) $\rightarrow$ \texttt{read} $\rightarrow$ \texttt{socket}(\texttt{AF\_INET}, \texttt{SOCK\_DGRAM}) $\rightarrow$ \texttt{sendto} \\
\cline{2-3}
 & System fingerprint + exfiltration &
\texttt{uname}/\texttt{sysinfo} $\rightarrow$ (\texttt{socket}) $\rightarrow$ \texttt{connect}(remote) $\rightarrow$ \texttt{sendto} \\
\cline{2-3}
 & Network discovery + exfiltration &
\texttt{socket}(\texttt{AF\_NETLINK}) $\rightarrow$ (\texttt{bind}) $\rightarrow$ \texttt{sendto}(route query) $\rightarrow$ \texttt{recvfrom} $\rightarrow$ \texttt{connect}(remote) $\rightarrow$ \texttt{sendto} \\
\cline{2-3}
 & Process enumeration + exfiltration &
\texttt{getdents64}(/proc) $\rightarrow$ \texttt{openat}(/proc/\textit{pid}/*, \texttt{O\_RDONLY}) $\rightarrow$ \texttt{read} $\rightarrow$ \texttt{connect}(remote) $\rightarrow$ \texttt{sendto} \\
\cline{2-3}
 & Command-based credential access &
\texttt{execve}(cat/grep/head) $\rightarrow$ \texttt{openat}(credential file) $\rightarrow$ \texttt{read} \\
\cline{2-3}
 & Keylogging + exfiltration &
\texttt{openat}(/dev/input/event*, \texttt{O\_RDONLY}) $\rightarrow$ \texttt{read} $\rightarrow$ \texttt{connect}(remote) $\rightarrow$ \texttt{sendto} \\
\cline{2-3}
 & Credential access during execution &
\texttt{openat}(credential file, \texttt{O\_RDONLY}) $\rightarrow$ \texttt{read} \\
\hline

\multirow{8}{*}{AB2}
 & Cron job installation &
\texttt{openat}(/etc/crontab or cron.d/*, \texttt{O\_WRONLY}) $\rightarrow$ \texttt{write} \\
\cline{2-3}
 & Systemd service creation &
\texttt{openat}(systemd unit path, \texttt{O\_WRONLY}) $\rightarrow$ \texttt{write} $\rightarrow$ (\texttt{symlink} to target.wants) $\rightarrow$ (\texttt{execve}(systemctl)) \\
\cline{2-3}
 & SSH authorized\_keys injection &
(\texttt{mkdir}(.ssh)) $\rightarrow$ \texttt{openat}(authorized\_keys, \texttt{O\_WRONLY}) $\rightarrow$ \texttt{write} $\rightarrow$ (\texttt{chmod}) \\
\cline{2-3}
 & Shell profile modification &
\texttt{openat}(.bashrc/.profile/.zshrc, \texttt{O\_WRONLY}) $\rightarrow$ \texttt{write} \\
\cline{2-3}
 & Dynamic linker hijack &
\texttt{openat}(/etc/ld.so.preload or ld.so.conf, \texttt{O\_WRONLY}) $\rightarrow$ \texttt{write} \\
\cline{2-3}
 & Binary replacement &
\texttt{openat}(/usr/bin/* or /usr/lib/*, \texttt{O\_WRONLY}) $\rightarrow$ \texttt{write} $\rightarrow$ (\texttt{chmod}) \\
\cline{2-3}
 & User account creation &
\texttt{openat}(/etc/passwd or /etc/shadow, \texttt{O\_WRONLY}) $\rightarrow$ \texttt{write} \\
\cline{2-3}
 & Python module shadowing &
\texttt{openat}(torch.py or numpy.py, \texttt{O\_WRONLY}) $\rightarrow$ \texttt{write} \\
\hline

\multirow{8}{*}{AB3}
 & System/log file deletion &
\texttt{unlink}(/var/log/* or /etc/* or /home/*) \\
\cline{2-3}
 & System file modification &
\texttt{openat}(/etc/* or /var/log/*, \texttt{O\_WRONLY}) $\rightarrow$ \texttt{write} \\
\cline{2-3}
 & Permission modification &
\texttt{chmod}/\texttt{fchmod}(system file) \\
\cline{2-3}
 & Symlink attack &
\texttt{unlink}(sensitive file) $\rightarrow$ \texttt{symlink} or \texttt{link} \\
\cline{2-3}
 & Firewall manipulation (Netlink) &
\texttt{socket}(\texttt{AF\_NETLINK}, \texttt{NETFILTER}) $\rightarrow$ \texttt{sendto} \\
\cline{2-3}
 & Process name masquerading &
\texttt{connect}(remote) $\rightarrow$ \texttt{prctl}(\texttt{PR\_SET\_NAME}) \\
\cline{2-3}
 & Temporary script staging &
\texttt{openat}(/tmp/*.sh or *.py, \texttt{O\_WRONLY}) $\rightarrow$ \texttt{write} \\
\cline{2-3}
 & Syslog message injection &
\texttt{socket}(\texttt{AF\_UNIX}, \texttt{SOCK\_DGRAM}) $\rightarrow$ \texttt{sendto} \\
\hline

\multirow{6}{*}{AB4}
 & Shell interpreter spawning &
\texttt{execve}(/bin/sh or /bin/bash) \\
\cline{2-3}
 & Script staging and execution &
\texttt{openat}(/tmp/script, \texttt{O\_WRONLY}) $\rightarrow$ \texttt{write} $\rightarrow$ (\texttt{chmod}) $\rightarrow$ \texttt{execve} \\
\cline{2-3}
 & Download, stage, and execute &
\texttt{connect}(remote) $\rightarrow$ \texttt{recvfrom} $\rightarrow$ \texttt{openat}(\texttt{O\_WRONLY}) $\rightarrow$ \texttt{write} $\rightarrow$ (\texttt{chmod}) $\rightarrow$ \texttt{execve} \\
\cline{2-3}
 & Fileless execution via memfd\_create &
\texttt{memfd\_create} $\rightarrow$ \texttt{write} $\rightarrow$ \texttt{execve} \\
\cline{2-3}
 & Process injection via ptrace &
\texttt{ptrace}(\texttt{PTRACE\_ATTACH}) $\rightarrow$ \texttt{ptrace}(\texttt{PTRACE\_POKETEXT}) $\rightarrow$ \texttt{ptrace}(\texttt{PTRACE\_DETACH}) \\
\cline{2-3}
 & Suspicious subprocess via pipe+fork &
\texttt{pipe2} $\rightarrow$ \texttt{clone} $\rightarrow$ \texttt{dup2} $\rightarrow$ \texttt{execve} \\
\hline

\multirow{3}{*}{AB5}
 & Miner execution with pool communication &
(\texttt{uname}/\texttt{sysinfo}) $\rightarrow$ \texttt{execve}(miner binary) $\rightarrow$ \texttt{connect}(mining pool port) $\rightarrow$ (\texttt{recvfrom}) $\rightarrow$ \texttt{sendto} \\
\cline{2-3}
 & Fileless miner execution &
\texttt{memfd\_create} $\rightarrow$ \texttt{write} $\rightarrow$ \texttt{execve} $\rightarrow$ \texttt{connect}(mining pool port) $\rightarrow$ (\texttt{recvfrom}) $\rightarrow$ \texttt{sendto} \\
\cline{2-3}
 & Miner deployment and execution &
\texttt{openat}(miner path, \texttt{O\_WRONLY}) $\rightarrow$ \texttt{write} $\rightarrow$ (\texttt{chmod}) $\rightarrow$ \texttt{execve}(same path) \\
\hline

\multirow{6}{*}{AB6}
 & Reverse shell &
\texttt{socket} $\rightarrow$ \texttt{connect}(remote) $\rightarrow$ \texttt{dup2}($\rightarrow$0) $\rightarrow$ \texttt{dup2}($\rightarrow$1) $\rightarrow$ \texttt{dup2}($\rightarrow$2) $\rightarrow$ \texttt{execve}(shell) \\
\cline{2-3}
 & Bind shell &
\texttt{socket} $\rightarrow$ \texttt{bind} $\rightarrow$ \texttt{listen} $\rightarrow$ \texttt{accept} $\rightarrow$ \texttt{dup2} $\rightarrow$ \texttt{execve}(shell) \\
\cline{2-3}
 & Pipe-based shell &
\texttt{pipe2} $\rightarrow$ \texttt{clone} $\rightarrow$ \texttt{dup2} $\rightarrow$ \texttt{execve}(shell) \\
\cline{2-3}
 & Remote command execution &
\texttt{connect}(remote) $\rightarrow$ \texttt{recvfrom}(command) $\rightarrow$ \texttt{execve} $\rightarrow$ (\texttt{sendto}(result)) \\
\cline{2-3}
 & Proxy/port forwarding &
\texttt{socket} $\rightarrow$ \texttt{bind} $\rightarrow$ \texttt{listen} $\rightarrow$ \texttt{accept} $\rightarrow$ \texttt{connect}(remote) \\
\cline{2-3}
 & Lateral movement via SSH &
\texttt{execve}(ssh) $\rightarrow$ \texttt{connect}(remote:22) \\
\hline

\multirow{4}{*}{AB7}
 & In-place encryption &
key-generation indicator $\rightarrow$ \texttt{getdents64}(user directory) $\rightarrow$ \texttt{openat} $\rightarrow$ \texttt{read} $\rightarrow$ \texttt{write} $\rightarrow$ (\texttt{rename}) \\
\cline{2-3}
 & Copy-and-delete encryption &
\texttt{getdents64} $\rightarrow$ \texttt{openat}(original, \texttt{O\_RDONLY}) $\rightarrow$ \texttt{read} $\rightarrow$ \texttt{openat}(.enc, \texttt{O\_WRONLY}) $\rightarrow$ \texttt{write} $\rightarrow$ \texttt{unlink}(original) \\
\cline{2-3}
 & Encryption key exfiltration &
key-generation indicator $\rightarrow$ \texttt{connect}(remote) $\rightarrow$ \texttt{sendto} \\
\cline{2-3}
 & Ransom note creation &
\texttt{openat}(README or RANSOM*, \texttt{O\_WRONLY}) $\rightarrow$ \texttt{write} \\
\hline

\multirow{4}{*}{AB8}
 & Direct privilege elevation &
\texttt{setuid}(0) or \texttt{setresuid}(0,0,0) \\
\cline{2-3}
 & Execute then elevate &
\texttt{execve}(binary) $\rightarrow$ \texttt{setuid}(0)/\texttt{setresuid}(0,0,0) \\
\cline{2-3}
 & Kernel module loading &
\texttt{init\_module}/\texttt{finit\_module}, or \texttt{execve}(insmod/modprobe) \\
\cline{2-3}
 & Namespace escape &
\texttt{setns}(namespace fd), or \texttt{execve} $\rightarrow$ \texttt{unshare}/\texttt{setns} \\
\hline
\end{tabular}

\end{table*}

\begin{table*}[!htbp]
\centering
\fontsize{9.5}{11}\selectfont
\caption{Composition of the \textsc{ModelShield} runtime environment pool.}
\label{env-pool}
\renewcommand{\arraystretch}{1.3}
\begin{tabular}{p{3.5cm}|p{\dimexpr\textwidth-3.5cm-4\tabcolsep-\arrayrulewidth\relax}}
\hline
\textbf{Environment} & \textbf{Installed Packages} \\
\hline
PyTorch (current) &
\texttt{torch}~2.4.0, \texttt{transformers}~4.44.0, \texttt{tokenizers}, \texttt{sentencepiece}, \texttt{dill}, \texttt{cloudpickle}, \texttt{huggingface\_hub}, \texttt{safetensors}, \texttt{numpy} \\
\hline
PyTorch (legacy) &
\texttt{torch}~1.13.1, \texttt{transformers}~4.30.0, \texttt{dill}, \texttt{cloudpickle}, \texttt{huggingface\_hub}, \texttt{safetensors}, \texttt{numpy} \\
\hline
TensorFlow / Keras &
\texttt{tensorflow}~2.17.0, \texttt{keras}~3.5.0, \texttt{torch}~(CPU), \texttt{huggingface\_hub}, \texttt{safetensors}, \texttt{numpy} \\
\hline
Classical ML &
\texttt{scikit-learn}~1.5.1, \texttt{joblib}~1.4.2, \texttt{xgboost}~2.1.1, \texttt{lightgbm}~4.5.0, \texttt{catboost}~1.2.7, \texttt{pandas}, \texttt{sympy}, \texttt{pypmml}~1.5.8, \texttt{py4j}, \texttt{torch}~(CPU), \texttt{dill}, \texttt{cloudpickle}, \texttt{huggingface\_hub}, \texttt{safetensors}, \texttt{numpy} \\
\hline
ONNX Runtime &
\texttt{onnxruntime}~1.18.1, \texttt{onnx}~1.16.2, \texttt{torch}~(CPU), \texttt{huggingface\_hub}, \texttt{safetensors}, \texttt{numpy} \\
\hline
\end{tabular}
\end{table*}
\begin{table*}[!htb]
\centering
\fontsize{10}{11}\selectfont
\caption{\textsc{ModelShield} detection results on the four TensorAbuse attacks~\cite{zhu2025my}.}
\label{table_tensorabuse_eval}
\renewcommand{\arraystretch}{1.3}
\setlength{\tabcolsep}{4pt}
\resizebox{\textwidth}{!}{%
\begin{tabular}{p{2.3cm}|p{0.6cm}|p{3.1cm}|p{8.6cm}|p{1.4cm}|p{1.4cm}}
\hline
\textbf{Attack} & \textbf{AB} & \textbf{Main Behavior} & \textbf{Runtime Evidence} & \textbf{Phase} & \textbf{MITRE} \\
\hline
Code Execution &
AB3 &
Temporary shell script staging &
\texttt{openat}$\rightarrow$\texttt{/tmp/exploit.sh} &
Execution &
T1059.004 \\
\hline

Shell Access &
AB2 &
SSH \texttt{authorized\_keys} injection &
\texttt{openat}$\rightarrow$\texttt{write}$\rightarrow$\texttt{.ssh/authorized\_keys.\allowbreak tempstate*} &
Execution &
T1098.004 \\
\cline{2-6}
&
AB6 &
Outbound connection to attacker &
\texttt{socket(AF\_INET6)}$\rightarrow$\texttt{connect(192.168.29.129\allowbreak:1236)} &
Execution &
T1071  \\
\hline
IP Exposure &
AB1 &
Hardware information theft and exfiltration &
\texttt{openat(/proc/cpuinfo)}$\rightarrow$\texttt{read}$\rightarrow$\texttt{connect(\allowbreak192.168.29.129:1236)}$\rightarrow$\texttt{sendmsg} &
Execution &
T1082, T1041 \\
\cline{2-6}
&
AB1 &
System fingerprinting and exfiltration &
\texttt{uname}$\rightarrow$\texttt{connect(192.168.29.129:1236) \allowbreak}$\rightarrow$\texttt{sendmsg} &
Execution &
T1082, T1041 \\
\cline{2-6}
&
AB6 &
Outbound connection to attacker &
\texttt{socket(AF\_INET6)}$\rightarrow$\texttt{connect(192.168.29.129\allowbreak:1236)} &
Execution &
T1071  \\
\hline
File Leak &
AB1 &
Hardware information theft and exfiltration &
\texttt{openat(/proc/\allowbreak cpuinfo)}$\rightarrow$\texttt{read}$\rightarrow$\newline
\texttt{connect(192.168.29.129:\allowbreak1236)}$\rightarrow$\texttt{sendmsg} &
Execution &
T1082,\newline T1041 \\
\cline{2-6}
&
AB1 &
System fingerprinting and exfiltration &
\texttt{uname}$\rightarrow$
\texttt{connect(192.168.29.129:\allowbreak1236)}$\rightarrow$\texttt{sendmsg} &
Execution &
T1082,\newline T1041 \\
\cline{2-6}
&
AB1 &
SSH key theft and exfiltration &
\texttt{openat(\$HOME/\allowbreak .ssh/\allowbreak id\_rsa)}$\rightarrow$
\texttt{pread64}$\rightarrow$\newline
\texttt{connect(192.168.29.129:\allowbreak1236)}$\rightarrow$\texttt{sendmsg} &
Execution &
T1552.004,\newline T1041 \\
\cline{2-6}
&
AB1 &
SSH private-key access &
\texttt{openat(\$HOME/\allowbreak .ssh/\allowbreak id\_rsa)}$\rightarrow$
\texttt{pread64} &
Execution &
T1552.004 \\
\cline{2-6}
&
AB6 &
Outbound connection to attacker &
\texttt{socket(AF\_INET6)}$\rightarrow$
\texttt{connect(192.168.29.129:\allowbreak1236)} &
Execution &
T1071 \\
\hline
\end{tabular}
}\\[4pt]
\begin{minipage}{0.96\textwidth}
\footnotesize
\textit{Note.} The endpoint \texttt{192.168.29.129} is the local IP address used in our experiment to represent the attacker-controlled server.
\end{minipage}
\end{table*}

\begin{table*}[!htbp]
\centering
\fontsize{14}{19}\selectfont
\caption{Summary of 41 confirmed malicious models detected by \textsc{ModelShield} on Hugging Face.}
\label{malicious_models_full}
\renewcommand{\arraystretch}{1.2}
\setlength{\tabcolsep}{4pt}
\resizebox{\textwidth}{!}{%
\begin{tabular}{r|l|l|r|r|c|c|c|l}
\hline
\multirow{2}{*}{\textbf{\#}} & \multirow{2}{*}{\textbf{Author/Model ID}} & \multirow{2}{*}{\textbf{Format}} & \multirow{2}{*}{\makecell{\textbf{Age at}\\\textbf{Detection}}} & \multirow{2}{*}{\textbf{\# Downloads}} & \multirow{2}{*}{\makecell{\textbf{Current}\\\textbf{Status}}} & \multirow{2}{*}{\makecell{\textbf{Detected by HF}\\\textbf{at Scan Time}}} & \multirow{2}{*}{\textbf{Phase}} & \multirow{2}{*}{\textbf{Behavior}} \\
 & & & & & & & & \\
\hline
1  & LQCTestingSpace/lqc\_model                                & PyTorch       & 3\,d   & 108  & R & N & E & AB4, AB6 \\
2  & mathiasthemaster/tabular-playground                       & Scikit-learn  & 1\,d   & 0    & R & N & E & AB4 \\
3  & 5oo6w32cu/test-model                                      & PyTorch       & 11\,h  & 43   & A & N & L & AB4, AB6 \\
4  & askj00bs/modelscan-test-1                                 & Joblib        & 11\,d  & 0    & A & Y & L & AB4 \\
5  & cw-fulmark/mlflow-dspy-bypass-rce                         & Pickle        & 16\,d  & 0    & R & N & L & AB4 \\
6  & jkhjkll/security-test-xss                                 & Pickle        & 3\,d   & 0    & A & N & E & AB4 \\
7  & m-j-contrary/transformers-bypass-poc                      & PyTorch       & 16\,d  & 23   & A & N & L & AB4 \\
8  & Deneimaiegw/modelscan-npy-bypass-poc                      & NumPy (.npy)  & 1\,d   & 0    & A & N & E & AB4 \\
9  & hacnho/pickle-scanner-bypass-syslog-syslog                & Pickle        & 1\,d   & 0    & R & N & L & AB3 \\
10 & hacnho/pickle-scanner-bypass-compileall-compile-file      & Pickle        & 1\,d   & 0    & R & N & L & AB3 \\
11 & hacnho/pickle-scanner-bypass-imaplib-imap4                & Pickle        & 1\,d   & 0    & R & N & L & AB6 \\
12 & hacnho/pickle-scanner-bypass-nntplib-nntp                 & Pickle        & 1\,d   & 0    & R & N & L & AB6 \\
13 & hacnho/pickle-scanner-bypass-py-compile-compile           & Pickle        & 1\,d   & 0    & R & N & L & AB3 \\
14 & hacnho/pickle-scanner-bypass-torch-multiprocessing-spawn  & Pickle        & 1\,d   & 0    & R & N & L & AB4 \\
15 & hacnho/poc-joblib-compressed-bypass                       & Joblib        & 1\,d   & 0    & R & N & L & AB4 \\
16 & hacnho/pickle-scanner-bypass-poplib-pop3                  & Pickle        & 1\,d   & 0    & R & N & L & AB6 \\
17 & hacnho/pickle-scanner-bypass-smtplib-smtp                 & Pickle        & 1\,d   & 0    & R & N & L & AB6 \\
18 & manja316/modelscan-bypass-timeit                          & Pickle        & 6\,h   & 0    & A & N & E & AB4 \\
19 & manja316/modelscan-bypass-timeit-repeat                   & Pickle        & 6\,h   & 0    & A & N & E & AB4 \\
20 & manja316/modelscan-bypass-pycompile                       & Pickle        & 17\,h  & 0    & A & N & L & AB3 \\
21 & manja316/modelscan-bypass-ssl-ssrf                        & Pickle        & 17\,h  & 0    & A & N & L & AB6 \\
22 & mscgo77/modelscan-bypass-ftplib                           & Joblib        & 5\,d   & 0    & A & N & L & AB6 \\
23 & mscgo77/modelscan-bypass-logging-config                   & Joblib        & 6\,d   & 0    & A & N & E & AB4 \\
24 & mscgo77/npy-scanner-bypass-poc                            & NumPy (.npy/.npz) & 23\,d & 0  & A & N & L & AB4, AB6 \\
25 & rajasingh010/keras-saveable-rce                           & Pickle        & 13\,d  & 0    & A & N & E & AB4, AB6 \\
26 & rajasingh010/vllm-torch-rce                               & PyTorch       & 13\,d  & 0    & A & N & E & AB4, AB6 \\
27 & Skyflocka/dill-importlib-rce-poc                          & Dill          & 19\,d  & 0    & A & N & E & AB4 \\
28 & Skyflocka/joblib-importlib-rce-poc                        & Joblib        & 19\,d  & 0    & A & N & E & AB4 \\
29 & Skyflocka/numpy-npy-importlib-rce-poc                     & NumPy (.npy)  & 20\,d  & 0    & A & N & E & AB4 \\
30 & Skyflocka/numpy-npz-modelscan-bypass-poc                  & NumPy (.npz)  & 20\,d  & 0    & A & N & E & AB4 \\
31 & Skyflocka/pickle-importlib-rce-poc                        & Pickle        & 19\,d  & 0    & A & N & E & AB4 \\
32 & Skyflocka/pytorch-importlib-rce-poc                       & PyTorch       & 19\,d  & 0    & A & N & E & AB4 \\
33 & TheAuroraAi/modelscan-denylist-gap-poc                    & Pickle        & 11\,d  & 0    & A & P & L & AB6 \\
34 & TheAuroraAi/profile-run-bypass-poc                        & Pickle + PyTorch & 12\,d & 0  & A & N & E & AB4 \\
35 & thisisauser101/picklescan-bypass-poc                      & Pickle        & 17\,d  & 0    & A & N & E & AB4 \\
36 & vellaveto/joblib-scanner-bypass-poc                       & Joblib        & 13\,d  & 192  & A & N & E & AB4, AB6 \\
37 & vellaveto/mfv-pickle-modelscan-bypass                     & PyTorch       & 13\,d  & 27   & A & Y & E & AB4 \\
38 & jorseckc/hazard                                           & Joblib        & 5\,d   & 0    & A & Y & L & AB4 \\
39 & cgruzicka/test-lora-weights                               & PyTorch       & 3\,d   & 0    & A & Y & L & AB1, AB4, AB6 \\
40 & SantanamoDon/huggingface-transformers-poc                 & PyTorch       & 18\,d  & 0    & A & Y & L & AB4 \\
41 & abdellahennajari/totally-harmless-model                   & PyTorch       & 20\,d  & 31   & A & Y & L & AB4, AB8 \\
\hline
\end{tabular}
}
\\[4pt]
\begin{minipage}{0.95\textwidth}
\footnotesize
Age at Detection: time between the repository's first commit and detection by \textsc{ModelShield}, reported in days (d) or hours (h). Current Status: A = Active (still hosted as of September 2026), R = Removed. Detected by Hugging Face at Scan Time: whether Hugging Face's malware scanner flagged the model at the time of our scan (Y = Yes, N = No, P = Partial, i.e., some files flagged). Phase: L = Loading, E = Execution. Behavior classes are defined in Table~\ref{attack_taxonomy}. Download counts are cumulative totals recorded at detection time.
\end{minipage}

\end{table*}

\end{document}